\documentclass[%
reprint,
superscriptaddress,
 amsmath,amssymb,
 aps,
 pra,
floatfix,
]{revtex4-2}
\newcommand{\angstrom}{ \text{\normalfont\AA}  }

\usepackage{graphicx}
\usepackage{dcolumn}
\usepackage{bm}

\begin{document}


\title{Real-Space Inversion and Super-Resolution of Ultrafast Scatterings}

\author{Adi Natan}
 \email{natan@stanford.edu}
\affiliation{Stanford PULSE Institute, SLAC National Accelerator Laboratory 2575 Sand Hill Road, Menlo Park, CA 94025}

\date{\today}

\begin{abstract}
Ultrafast scattering using X-rays or electrons is an emerging method to obtain structure dynamics at the atomic length and time scales.  However, directly resolving in real-space atomic motions is inherently limited by the finite detector range and the probe energy. As a result, the time-resolved signal interpretation is mostly done in reciprocal space and relies on modeling and simulations of specific structures and processes. Here, we introduce a model-free approach to directly resolve scattering signals in real space, surpassing the diffraction limit, using scattering kernels and signal priors that naturally arise from the measurement constraints. We demonstrate the approach on simulated and experimental data, recover multiple atomic motions at sub-$\angstrom$ngstrom resolutions, and discuss the recovery accuracy and resolution limits vs signal fidelity.  The approach offers a robust path to obtain high-resolution real-space information of atomic-scale structure dynamics using current time-resolved X-ray or electron scattering sources.
\end{abstract}

\maketitle

\section{Introduction}

Probing the structure and dynamics of systems of increased complexity with atomic resolution, ranging from small molecules in the gas phase to solute-solvent systems and disordered materials, is at the forefront of experimental research. Ultrafast scattering (unlike diffraction) usually refers to the diffuse scattering from atomic-size charge densities or $\angstrom$ngstrom-range structural correlations, using X-rays or electron pulses whose durations are below a picosecond.  X-ray free-electron lasers (XFELs) and relativistic or high-energy electrons are sources of such pulses and have become powerful tools to study ultrafast atomic and molecular dynamics in chemical and solid-state systems at the $\angstrom$ngstrom and femtosecond scales.   

In typical experiments that employ ultrafast scattering,  an ultrashort optical "pump" pulse is used to photo-excite the system under study, and the probing is done at some time delay via an ultrashort X-ray or electron pulse that scatters from the sample. The time-delayed signal is usually subtracted from the signal of the unexcited sample, to allow tracing changes in signal positions and infer structural dynamics. For example, this method was recently applied to probe molecules in the gas phase following single and multi-photon excitations and chemical reactions \cite{kupper2014x,PhysRevLett.114.255501,PhysRevLett.117.153003,bucksbaum2020characterizing,natan2021resolving,kierspel2020x,wolf2019photochemical,champenois2021conformer,yang2020simultaneous}, as well as structural changes of molecules in solution environments  \cite{ihee2010ultrafast,kim2015direct,PhysRevLett.117.013002,van2016atomistic,Chollet,PhysRevLett.122.063001,panman2020observing}.  
 
Directly obtaining real-space time-resolved information by inversion of the scattering pattern is often not possible in these studies because of the limited available range of the scattering vector, severely restricting the spatial resolution. 
To address this issue, the methods that were developed to interpret the scattering signals were system dependent and relied on calculating and attaining trajectory statistics. These methods were mostly restricted to the reciprocal q-space \cite{minitti2015imaging,stankus2019ultrafast}, and often limited to a particular reaction pathway.  Complex atomic motions of general polyatomic systems that take place simultaneously and involve multiple pathways at sub-$\angstrom$ngstrom distances are theoretically challenging to model and were not directly resolved in experiments.

Here we demonstrate a robust and model-free deconvolution approach to overcome the limitations that are imposed by a typical inversion procedure, extending super-resolution methods that have recently transformed optical microscopy and biological imaging \cite{hell1994breaking,betzig2006imaging,sigal2018visualizing,phillips2020cryosim,sroda2020sofism}, to applications where only restricted scattering information is present, with no access to high spatial frequencies, multiple scattering, or single emitters.  
Our approach does not require a pool of calculated structures, or assume any structure at all, making it attractive for capturing general motions that may contain density distributions beyond atomic positions, such as wavepacket motion, as well as differential signals where an undetermined subset of the system contributes to the measured motion.

We show how real-space recovery of multiple and complex motions of general polyatomic systems can be obtained by leveraging the information and constraints that naturally arise from the measurement configuration and analysis procedure. We derive a formalism to express the way a set of scattering kernels, which are effectively position-dependent point spread functions in ultrafast scattering experiments, can be used to capture general distortions in the inversion procedure. 
We then use regularization and convex optimization to recover and super-resolve real-space atomic motions as a sparse or smooth solution of a dictionary of scattering kernels. While the approach is applicable for X-ray, electron, and neutron scattering, for brevity we'll discuss mostly the case of X-ray scattering.

\section{Method}

To describe $\angstrom$ngstrom and femtosecond scale dynamics of general molecular systems, it is not valid to assume sample periodicity. As a result, the scattering patterns are often broad and diffuse, capturing short-range spatial correlations.  Diffuse X-ray scattering of time-evolving charge densities is generally inelastic, but under typical experimental conditions, the general electronic scattering operator can be replaced with its elastic expression
\cite{Lorenz_Moller_Henriksen_2010,simmermacher2019electronic}.  In experiments, the signal is usually integrated over angle for improved fidelity, and considering the isotropic signal was shown to contain all the nuclear and electronic structural evolution of the charge density \cite{lorenz2010interpretation}. For simplicity, we'll introduce the approach using the isotropic scattering signal  given by the Debye equation \cite{debye1915zerstreuung,dohn2015calculation}:

\begin{multline}
S_0(q,\tau) = \sum_{a} f_a(q)^2+ \\ \sum_a \sum_{b\ne a} f_a^*(q) f_b(q) 4 \pi\int_0^{\infty}  d R \, R^2 \rho_{ab}(R,\tau) j_0(qR),
\label{eq:S0}
\end{multline}

where $S_0(q,\tau)$ is the isotropic diffuse scattering signal at time delay $\tau$, $q{=}4\pi \sin(\theta_q{/}2){/}\lambda$ is the scattering vector magnitude, with the scattering angle $\theta_q$ and wavelength $\lambda$, the double sum is over all atom pairs, $f_i(q)$ is the atomic form factor of the $i^{th}$ atom,   $\rho_{ab}(R,\tau)$ is the pair charge density, and $j_0(qR){=}sin(qR){/}qR$ is the zeroth order spherical Bessel function of the first kind. 

For the case of ultrafast scattering, the scattering signal difference $\Delta S_0(q,\tau){=}S_0(q,\tau){-}S_0(q,0)$ is often analyzed,  and as a result, the  stationary atomic form factors term $\sum_a f_a(q)^2$ in Eq. \ref{eq:S0} cancels. Without loss of generality, we'll therefore consider the scattering difference signal to demonstrate the approach to recover the isotropic real-space difference pair density  $\Delta \rho_{ab}$.   The Debye equation is related to the real-space pair correlation function through a Fourier transform \cite{zernike1927beugung}. For isotropic signals, direct real-space inversion of Eq. \ref{eq:S0} can be  done for the case of a single  pair density contribution:
\begin{equation}
\Delta\rho_{ab}(R,\tau) {=}   \int_0^{\infty} dq \, q^2 \frac{ \Delta S_0(q,\tau) j_0(qR)}{f_a(q)f_b(q)}.
\label{eq:PD0}
\end{equation}

 The integrand here is often scaled with an exponential function  $e^{-kq^2}$  \cite{williamson1994ultrafast,kim2015direct} to serve as an effective experimental integration bound, however, this formalism does not capture various aspects of the measurement, such as the experimental configuration, sample thickness, detector truncation, and discretization. As a result, applying Eq. \ref{eq:PD0} to  experimental ultrafast scattering signals will contain inversion artifacts in addition to the diffraction-limited resolution, especially around the few  $\angstrom$ngstrom length scale that is relevant for structural dynamics of atomic and molecular systems.

 \begin{figure}[hbt!]
	\includegraphics[width=0.5\textwidth]{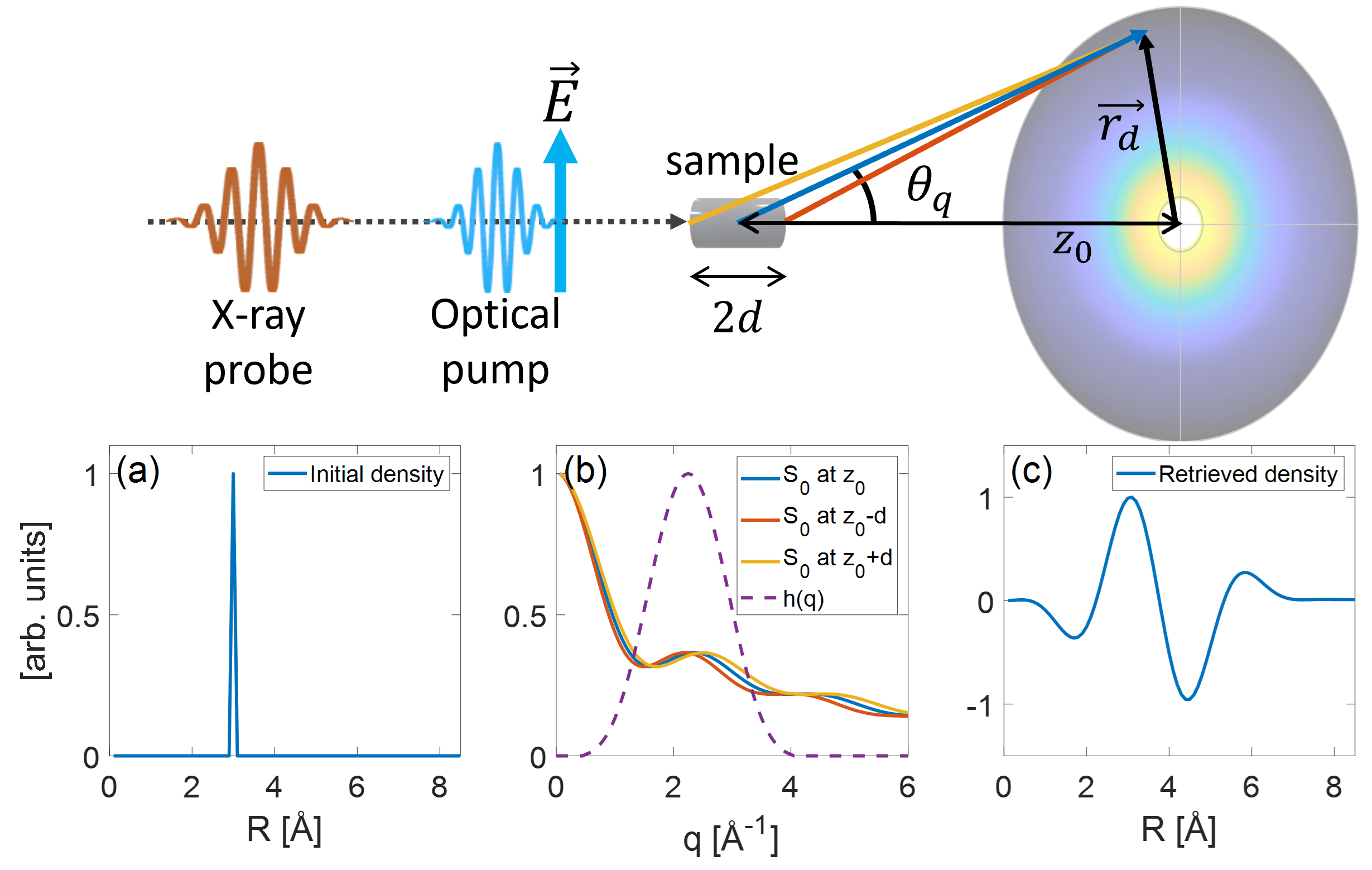} 
	\caption{(top) A sketch of a typical optical pump X-ray scattering probe experimental configuration, with sample thickness $2d$ located $z_0$ away from a finite detector. We illustrate how the scattering on the detector at a radius $r_d$ is the sum of different parts of the sample (depicted as yellow, blue, and red lines) as explained below.  (a) Using a Dirac delta at 3$\angstrom$ as the initial pair density, we model the scattering using sample thickness and distance from a detector similar to \cite{PhysRevLett.117.153003}. (b) The isotropic scattering curves $S_0(q)$  from the sample center at $z_0$ (blue) vs the sample edges at $z_0 {\pm} d$ (red and yellow) add to the same spot in the detector, creating a signal distortion. The  scattering is also truncated due to the finite detector dimensions and q-dependent absorption, modeled by the window function $h(q)$. (c) The inversion of the total scattering given the measurement constraints captures how the distortion propagates to real space, where the retrieved pair density fails to accurately describe the initial density.}
	\label{fig:nsk_steps}
\end{figure} 

To illustrate this, in Fig \ref{fig:nsk_steps} we show the inversion of a Dirac delta charge density of a single atom pair at 3$\angstrom$. We assume typical experimental conditions for gas phase scattering  similar to \cite{PhysRevLett.117.153003}, e.g., 9 keV photon energy,  sample thickness of 8 mm, located 60 mm from a finite detector, a limited q-range $0.5{<}q{<}4 \angstrom^{\text{-}1}$,  resulting in a  diffraction limited spatial resolution given by $2\pi{/}q_{max}{\simeq}1.57\angstrom$.  The inverted waveform in Fig \ref{fig:nsk_steps}c fails to accurately recover the original delta function position as it is distorted by the scattering integrated over the sample thickness and q-range truncation. This example provides us with a path to define a Natural Scattering Kernel (NSK), or the function that describes the distortion that a delta function charge density naturally undergoes in terms of measurement in $q$-space, and inversion analysis in real space. Because we can express any pair density as a weighted sum of delta functions in real space, we would like to obtain a dictionary of NSKs  that can explain experimental ultrafast scattering signals. We'll describe the considerations to obtain these kernels, and the way they can be used to recover and super-resolve real-space motions.

\subsection{Distortions in q-space}

The experimental configuration and finite detector may introduce signal distortions to the measured scattering pattern, as seen in Fig \ref{fig:nsk_steps}. We can express the relation between the scattering vector magnitude $q$, the distance of the sample from the detector $z$, and the radial position on the detector $r_d=\sqrt{x^2+y^2}$ by:
 \begin{equation}
q(z)= \frac{4\pi }{\sqrt 2 \lambda} \sqrt{1-\frac{z}{\sqrt{r_d^2+z^2}}}     ,
\label{eq:qrz}
\end{equation}
where the origin of the coordinates is located at the center of the detector, and we assume that the dimensions of the optical-X-ray pump-probe interaction area are much smaller than the sample thickness (typically microns vs millimeters). A general distortion of the scattering signal in q-space can be expressed by:

 \begin{equation}
 \tilde{S}_0(q,\tau){=} \int_{z_0{-}d}^{z_0{+}d}  dz
  \, h(q(z)) \, S_{0}(q(z),\tau)+b(q(z))    .
\label{eq:distortion}
\end{equation}
Where $\tilde{S}_0(q,\tau)$ is the scattering signal measured on the detector which is distorted by the measurement constraints, $S_0$ was defined in Eq.\ref{eq:S0}, and we assume a sample thickness of $2d$, at a distance of $z_0$ from the detector. We define $q{=}q(z_0)$ at the sample center, $b(q)$ is a possible additive background, and $h(q)$ is a window function that represents the detector's finite $q$-range as well as $q$-dependent signal absorption.  In experiments, the shape of $h(q)$ may also be affected by the sample density profile, propagation-related absorption, and detector-dependent attenuation related to the experimental setup design. These can be accounted for in calibration measurements, for example, in  \cite{budarz2016observation}. Here, we also assume that the measured signal was corrected for X-ray polarization, detector geometry and non-linear response \cite{van2020epix10k}.  In addition, in the analysis stage, a smooth modification function is often introduced to attenuate the signal cut-off at the limits of the detected q-range, such as the Slepian or the Lorch functions \cite{lorch1969neutron}. Here, the application of $h(q)$ accounts for such functions as well as all other contributions mentioned.

\subsection{Discretization}
Because the measured scattering signal is discretized by the detector at some resolution,  but the inversion  in Eq \ref{eq:PD0} is a continuous transform, we'd like to use a resampling and discretization scheme that maintains the transform's orthogonality and operational rules, such as discrete shift, multiplication, and convolution, between the $R$ and $q$ domains.  Using the Bessel-Fourier expansion  \cite{Baddour:15} and the standard Shannon sampling theorem \cite{shannon1949communication}, we define:

 \begin{align}
   \Bigl\{ q_{i} \Bigr\}_1^N &= \frac{j_{0i}}{j_{0N}}N \Delta q ,& \Bigl\{ R_{m} \Bigr\}_1^M &= \frac{j_{0m}}{M \Delta q}.
   \label{eq:discretization}
  \end{align}
Where we assume detector discretization $\Delta q$, with an experimental upper bound $q_{max}{=}N \Delta q$, and resample the scattering signal at $q_i$, given by $ j_{0i}$, the $i^{th}$ root of the spherical Bessel function $j_0$.  
The ratio $M{/}N$ is an effective super-resolution factor \cite{candes2006robust} that can be used to determine the final recovery resolution of noisy signals for a given signal-to-noise ratio (SNR).

\subsection{Inversion}
The distorted difference scattering signal is discretized $\Delta  \tilde{S}_0(q_i,\tau)$  and inverted using the resampling scheme we introduced in Eq. \ref{eq:discretization} by:

  \begin{equation}
 \Delta PD_0(R,\tau)  {=} \frac{[(M{-}1) \Delta q] ^2}{j_{0 M} } \sum_{i=1}^{M{-}1}  \bm{G}_{mi} \frac{  \Delta \tilde{S}_0(q_i,\tau) q_i }{ f_e(q_i)}   , 
 \label{eq:qtoR}
\end{equation}

where $\Delta PD_0$ is the discretized pair density difference,  $f_e(q_i) {=} \sum_{b{\ne} a} f_a(q_i) f_b(q_i)$ is an  effective form factor, and the transformation kernel is given by:

\begin{align}
   [\bm{G}]_{mi}   = 2\frac{j_0 \left( \frac{j_{0m}j_{0i} }{j_{0M}} \right) } {j_{0M} j_1^2(j_{0i})}  ,
 \label{eq:Gki}
\end{align}

where $j_1(j_{0i})$ is the first-order spherical Bessel function evaluated on the i$^{th}$ root of $j_0$ . The derivation of $\bm{G}$ is obtained in a similar fashion to the derivation done for the discrete Hankel transform \cite{Baddour:15}, adapted here for the case of spherical Bessel functions.  We note that $f_e$  is an approximation needed  in cases where the scattering signal is composed of several atom pair types with different form factors that are not separable in the analysis process, as only the total signal $S_0(q,\tau)$ is measured.  The approximation is justified because each $f_a(q)f_b(q)$ term is a featureless monotonic function in the truncated $q$-range, and so its real space inversion will mainly contribute to the amplitude at $0{<}R{<}\pi/q_{max}$. As a result, any inaccuracy using $f_e$ will be limited to an amplitude artifact in that range, because it is in the sub-$\angstrom$ngstrom range, which is smaller than typical bond lengths for typical  ultrafast scattering experiments.

To invert the total scattering signal $S_0(q_i,\tau)$  we can replace $\Delta  \tilde{S}_0(q_i,\tau)$ in Eq. \ref{eq:qtoR}  with  $\tilde{S}_0(q_i,\tau){-}\sum f_a(q_i)^2$. 

To clarify the steps above we'll consider the following example:
We solved a time-dependent Schr\"{o}dinger equation (TDSE) for diatomic Iodine with parameters similar to \cite{PhysRevLett.117.153003},  30 fs pulses at 520 nm, with 8$\%$ excitation fraction.  At a delay of $\tau{=}$150fs, the calculated density difference $\Delta \rho(R,\tau)$  
features a depleted ground state at 2.67$\angstrom$, with bound and dissociative excited states at 3.9$\angstrom$ and 4.7$\angstrom$ (Fig \ref{fig:dicts}a).   We simulated the scattering signal difference from this delay assuming the experimental conditions mentioned earlier  (8 mm sample width, 60 mm from a detector), with $\Delta q{=}0.1\angstrom^{\text{-}1}$, q-range of $0.5{<}q{<}4\angstrom^{\text{-}1}$ that is modeled as $h(q)$ by a Slepian window function ($N{=}40$ sampling points in q). The choice of $\Delta q$ magnitude was to capture details in real space limited by $R_{max} {=}   \pi{/}\Delta q {\simeq} 31.4 \angstrom$.

We applied the inversion step using real space discretization at a resolution of $\Delta R{\simeq} 0.157 \angstrom$ which ($1{/10}$ the diffraction limit for the q-range considered). The value of $\Delta R{\simeq}$  is  attained  by choosing $M{=}\text{ceil} (\pi/(\Delta R\Delta q) ){+}1$ sampling points, resulting with super-resolution factor $M{/}N {\simeq} 5$.   The inverted pair density difference $\Delta PD_0(R,\tau)$ in Fig \ref{fig:dicts}a cannot resolve the two excited states, as their separation ${\sim} 0.8 \angstrom$ is  below the diffraction limit (${\sim} 1.57\angstrom$). Furthermore, $\Delta PD_0(R,\tau)$  is distorted by the measurement constraints, resulting in a structure that has little to do with the initial charge density, as seen in the positive peak at 1$\angstrom$ and negative peak around 5.5$\angstrom$.

 
\subsection{Dictionary formation}
 
We'll implement the steps above: the q-space distortion, discretization, and inversion, for a Dirac delta pair density,  by substituting  $\rho_{ab}(R){=}\delta(R{-}R_m)$ in Eq.\ref{eq:S0}, as seen in Fig \ref{fig:nsk_steps}c for the case $R_m{=}3\angstrom$. The result will be the real-space distorted inversion of the delta function for position $R_m$,  expressed by:

  \begin{equation}
 NSK(R_m)  {=} \frac{ [(M{-}1) \Delta q] ^2   }{j_{0 M} } \sum_{i=1}^{M{-}1}  \bm{G}_{mi}   \tilde{j_0}(q_iR_m) q_i   , 
 \label{eq:NSK_deltaS}
\end{equation}
where, 

 \begin{equation}
\tilde{j}_0(q,R_m){=} \int_{z_0{-}d}^{z_0{+}d}  dz
  \, h(q(z)) \, j_{0}(q(z)R_m)    .
  \end{equation}

 For very thin samples where we can take $d{\to0}$ such as in the case of micron width liquid jets,  we can approximate $\tilde{j}_0(q,R_m){\simeq} j_0(q R_m) h(q)$.

The NSKs are system agnostic as the effective form factor $f_e$ that was used for the inversion of a general measured signal cancels for the case of the kernels, and the expression in Eq. \ref{eq:NSK_deltaS} simplifies to contain only the distorted $\tilde{j}_0(q,R_m)$.  We can now create a set of NSKs along the real space sampling given by Eq. \ref{eq:discretization}  to assemble a dictionary of the following form: 
\[    \bm{\mathcal{D}}{=} \begin{bmatrix}
\vline &  & \vline &    & \vline \\
NSK(R_1) & \cdots &  NSK(Rm)  & \cdots & NSK(R_M) \\
\vline &  & \vline &   & \vline 
\end{bmatrix}. \]



\begin{figure}[hbt!]
	\includegraphics[width=0.5\textwidth]{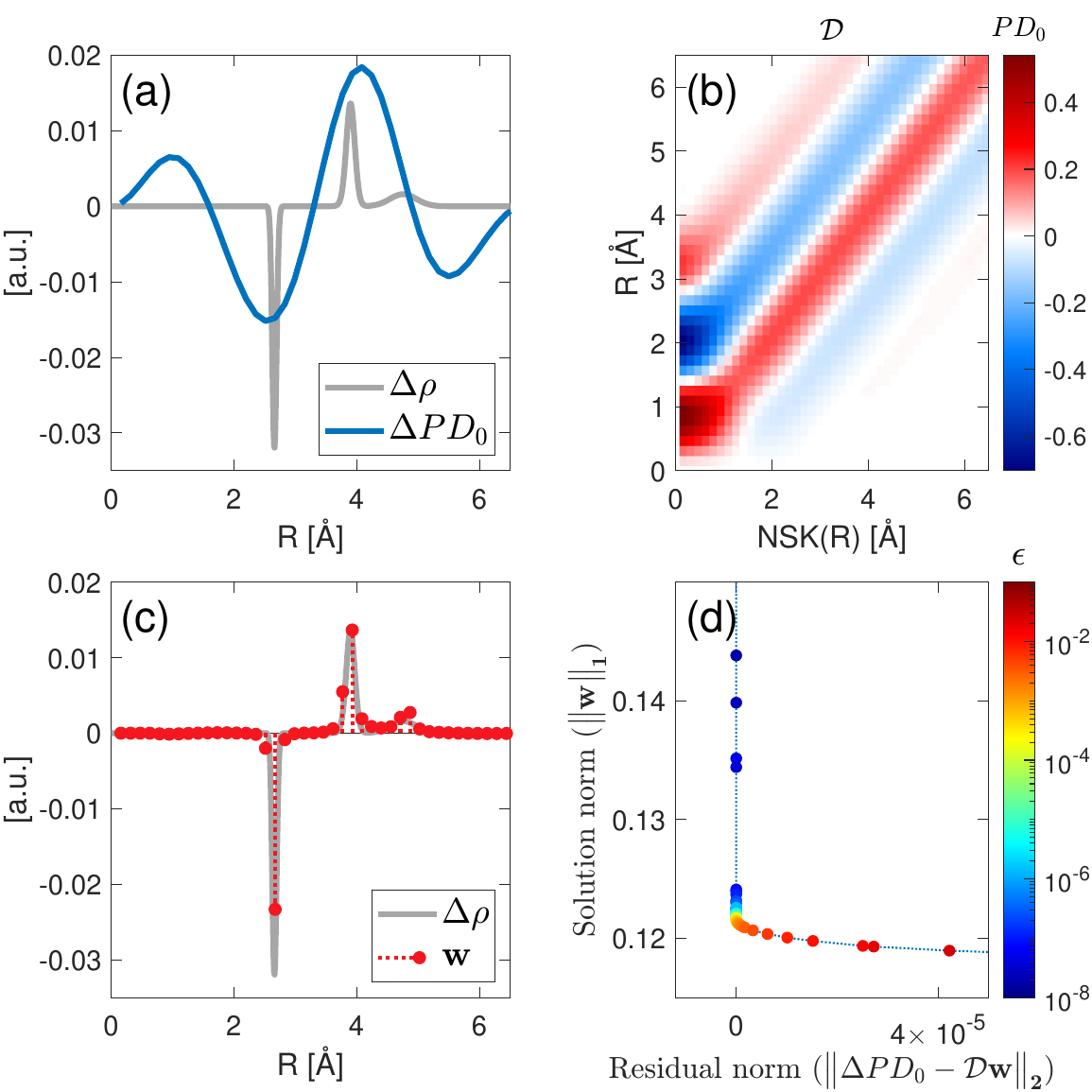} 
	\caption{(a) A simulated pair density difference $\Delta \rho$ of excited diatomic Iodine at a pump-probe delay of $\tau{=}150$fs is used to obtain the inverted pair density difference $\Delta PD_0$ (Eq \ref{eq:qtoR}), distorted by experimental conditions similar to \cite{PhysRevLett.117.153003}, where the truncated scattering range  $0.5{<}q{<}4\angstrom^{\text{-}1}$.  (b) Using the measurement constraints we form the NSK dictionary $\mathcal{D}$. We apply convex optimization for the  $\ell_1$  regularization case (Eq. \ref{eq:Regu}) and obtain (c) the NSKs weights vector $\bf{w}$ solution (red stem). The solution is in excellent agreement with the theoretical pair density, resolving peak positions and amplitudes  below the diffraction limit. (d) To obtain the  optimal value of the regularization parameter $\epsilon$ used in Eq. \ref{eq:Regu} we use the L-curve method by plotting the solution norm vs the residual norm for a logarithmic range in  $\epsilon$, using the value found at the corner of the L-shaped region (see text).} 
 \label{fig:dicts}  
	\end{figure}

In Fig \ref{fig:dicts}b we describe an NSK dictionary assuming the experimental conditions used in the example, similar to \cite{PhysRevLett.117.153003}.  The observed  position-dependent point spread function shape of the kernels is dominated by the shape of $h(q)$, which serves as a band-pass filter in q-space.

 \subsection{Deconvolution}
We'd like to use the NSKs dictionary we derived to explain the inversion signal that is distorted by the measurement constraints, assuming the linear model:
  
\begin{equation}
\Delta PD_0 = \bm{\mathcal{D}} \bf{w} ,
\label{eq:LinearModel}
\end{equation}
  
 Where we seek to estimate the weights vector $\bf{w}$ to recover the  pair density described by $\Delta \rho(R){=}\sum_m \text{w}_m \delta(R{-}R_m)$.  
 
 Naively, one would attempt to solve this  model using least-squares minimization: 
$\min_{\text{w}}     \big\|  \Delta PD_0 -\bm{\mathcal{D}}  \bf{w}  \big\|^2 $,  however, this approach is highly sensitive to noise and is generally unstable numerically when the problem is ill-posed.  A standard approach  for solving ill-posed inverse problems is to use a regularization framework of the form:

 \begin{equation}
 \min_{\text{w}}     \big\|  \Delta PD_0 -\bm{\mathcal{D}} \bf{w}  \big\|^2 \ + \epsilon \mathcal{R}( w )
 \label{eq:Regu}
\end{equation}

Where the regularizer $\mathcal{R}(\text{\bf{w}})$ is chosen according to some prior information of the measurement to promote solutions with preferable features such as sparsity or smoothness, and $\epsilon$  controls the magnitude of the regularization and can be estimated via the L-curve method using residual and solution norms or by cross-validation \cite{hansen1993use,arlot2010survey}.
For the examples discussed here, we'll implement the L-curve method.

Briefly, this curve is obtained when plotting the penalty term of the regularized solution norm $\mathcal{R}(\text{\bf{w}})$  vs the residual norm $\big\| \Delta PD_0 -\bm{\mathcal{D}} \bf{w} \big\|$$^2_2$ for a logarithmic range  of $\epsilon$ values, creating a characteristic L-shape, where $  \|\cdot \|_{2} $ is the Euclidean norm.

The $\epsilon$ value that corresponds to the corner of the L-curve, where the curvature is maximal, captures the best trade-off between minimizing the residual norm and the penalty term that captures the nature of the solution. An implementation of the L-curve method is shown in Fig \ref{fig:dicts}d for the example we discuss. 
  
For the regularizer $\mathcal{R}(\text{\bf{w}})$ we'll discuss two approaches,  promoting smooth or sparse solutions. A widely used  approach that addresses the numerical instabilities and produces low variance solutions is the Tikhonov or $\ell_{2}$ regularization \cite{golub1999tikhonov}, for which $\mathcal{R}(\text{\bf{w}}){=}\sum_m|\text{w}_m|^2$. Using this regularizer provides the closed-form linear solution: $ \bf{w}  {=} (\bm{\mathcal{D}}^T\bm{\mathcal{D}}{+}\epsilon \bm{1}    )^{-1} \bm{\mathcal{D}}^T$$\Delta PD_0$   that can be solved using singular value decomposition. This approach promotes stable and smooth solutions and increases the ability to predict their nature. However, it also often results in most of $\bf{w}$ elements having non-zero values, which may reduce the ability to resolve or explain weaker signals.

For the case where we assume that the solution for $\bf{w}$ is sparse, such that the number of sampling points in $R$ that can explain the distorted $\Delta PD_0$ is much smaller than $M$,  we can use the $\ell_1$ regularized least-squares model \cite{chen1994basis,tibshirani1996regression}, where $\mathcal{R}(\text{\bf{w}}){=}\sum_m |\text{w}_m|$. While this case doesn't have a closed-form solution, it can be solved using convex optimization, and lead to super-resolution. In recent years, several algorithmic approaches have evolved to solve this model, determine the optimal parameters \cite{boyd2011distributed,parikh2014proximal,candes2005l1}, and provide applications that transformed areas of research such as statistics, machine learning, and signal processing  \cite{lee2007efficient,mairal2009online,gregor2010learning,donoho2006compressed,candes2006robust,lustig2007sparse,duarte2008single,katz2009compressive,shechtman2010super,solomon2018sparsity,zhang2018tabletop,driver2020attosecond}. 

In Fig \ref{fig:dicts}c  we applied convex optimization for the $\ell_1$ regularization case (Eq.\ref{eq:Regu}) for the pair density difference and the NSK dictionary shown in Fig \ref{fig:dicts}a-b  using CVX, a package for specifying and solving convex programs \cite{grant2014cvx}. 
 
We find the optimal regularization parameter $\epsilon$ using the L-curve method (Fig \ref{fig:dicts}d). We note that the solution obtained is robust for a wide range  of $\epsilon$ values around the corner of the L-curve. The solution given by the weights vector \textbf{w} is in excellent agreement with the theoretical pair density difference,  super-resolving the excited states pair-density positions and amplitudes while eliminating the distortions in $\Delta PD_0$.

We note that the case presented here differs from some of the approaches referenced above that also used $\ell_1$ regularized least-squares. This is because the NSK dictionary  is highly correlated, each kernel is similar to its adjacent neighbors, and so the restricted isometry property  that is needed in compressed sensing schemes \cite{candes2006robust} is not satisfied here.

As a result, an additional restriction is introduced in the form of a minimum separation distance between adjacent NSKs, to avoid solution ambiguity below some $\delta R  {<} \nu{/}q_{max}$,  where $\nu$  usually depends on the signal's properties and fidelity  \cite{candes2013super,fernandez2016super,duval2015exact}. In the next section, We will discuss the conditions for super-resolution and find the minimum separation distance for noisy signals.

\section{Recovery accuracy for noisy signals and super-resolution} %

 \begin{figure}[hbt!]
	\includegraphics[width=0.5\textwidth]{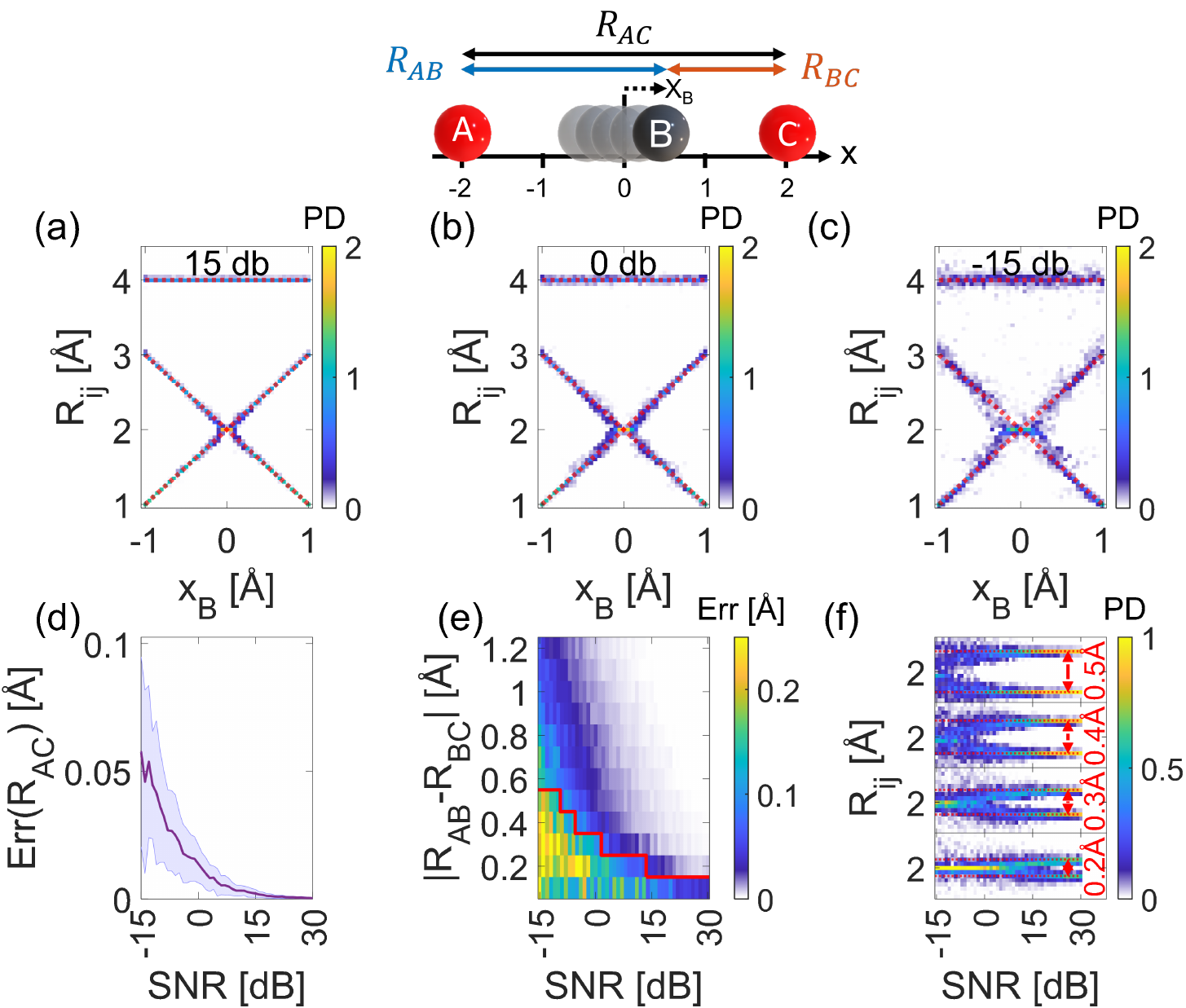} 
	\caption{Recovery accuracy under noise is tested for a 3-atom system (top), where two pair-distances  ($R_{AB}$, $R_{BC}$) approach each other below the diffraction limit. The scattering is simulated for each atom B position, truncated by $0.5{<}q{<}4 \angstrom^{\text{-}1}$, and inverted using NSKs ($\ell_1$ regularization). The diffraction limit for this case is $2\pi{/}q_{max}{\simeq}1.57$. The inversions for (a) 15, (b) 0, and (c) -15 dB SNRs are shown vs exact distances (dotted red), and $PD$ is normalized by the total charge density. The recovery error is proportional to the SNR, validating the stability of the approach. (d) For $R_{AC}$ distance, the average recovery error (solid) and standard deviation (shaded) are ${<}0.1 \angstrom$ across SNRs. The conditions for super-resolution and minimal separation emerge when the $R_{AB}$, $R_{BC}$ distances  approach each other. (e) The average recovery error (Err) is plotted as function of  pair separation $|R_{AB}{-}R_{BC}|$ and SNR, and used to define the minimal separation both distances are resolved (solid red). (f) The recovery in the super-resolution regime for different separations compared with the exact distances (dotted red) is limited by both SNR and the pair coalescence effect (see text). Here the the diffraction limit is surpassed by a factor $3.1{-}7.8$ for the range of -$15-30$ db.}
	\label{fig:noise}  
	\end{figure} 


 Super-resolution in the context of  scattering means to resolve at least two pair distances below the diffraction limit given by $2\pi/q_{max}$. The accuracy of the recovery method presented and the ability to achieve super-resolution depend both on the SNR, and the super-resolution factor.  To robustly super-resolve using NSKs and $\ell_1$ regularization, an additional minimum separation distance is introduced \cite{candes2013super}. To obtain its value and characterize the recovery accuracy we consider the case when two pair distances approach each other and focus on the recovery accuracy as a function of SNR.

We use a 3-atom model system, similar to CO$_2$  (Fig \ref{fig:noise}), where we fix the positions of the outer atoms (A, C) to $x{=}{\pm} 2\angstrom$ while changing the position of (B) the central atom (-$1\angstrom{<}x{<}1\angstrom$).  When atom $B$ is at $x{=}0$, the pair-distance $R_{AB}$ will merge with $R_{BC}$ at  $2 \angstrom$. We applied the same experimental configuration used in \cite{PhysRevLett.117.153003} and in the previous example, $q$-range $0.5{<}q{<}4\angstrom^{\text{-}1}$ with the discretization  $\Delta q{=}0.1 \angstrom^{\text{-}1}$,  and simulated the scattering signal for each of the distances. To test the recovery accuracy we added to the total scattering signal $q$-dependent additive white Gaussian noise, where each $q_i$ bin has sampling statistics proportional to the number of detector pixels contributing to it, assuming  the  array detector in \cite{van2020epix10k}.  
 
 We applied the NSK approach with sampling  $\Delta R{\simeq}0.05 \angstrom$, and tested the recovery for a range of single detector pixel SNRs (-15 to 30 dB), with 20 realizations to each noise level to obtain recovery statistics.
We find that the recovery accuracy for the $\ell_1$ regularization is proportional to the noise level, demonstrating the stability of the approach, and in agreement with theory \cite{candes2013super}. For the fixed $R_{AC}$ distance the recovery error is ${<}0.1 \angstrom$ across the SNRs (Fig \ref{fig:noise}d), demonstrating the accuracy of the approach for the case of well-separated pair distances (Fig \ref{fig:noise}c). For the case where the two pair distances $R_{AB}$, $R_{BC}$ approach each other, we obtain a minimal separation of $\delta R {\simeq} 0.35 \pm 0.15 \angstrom$ that allows robust recovery below the diffraction limit ($\simeq 1.57 \angstrom$) for the SNR range used (Fig \ref{fig:noise}e). The minimal separation distance is derived from the recovery error statistics, where we observe that the recovery error is both due to noise and a coalescence effect where both distances drift toward the joint center of mass (Fig \ref{fig:noise}f).  The coalescence effect is unique to the super-resolution case,  it has quadratic dependence in $M{/}N$ the super-resolution factor \cite{candes2013super}, and takes place when adjacent NSKs explain effectively the same noisy distorted measurement. As a result, the $\ell_1$ regularization that acts as a sparsity prior to the recovery will instead promote a single NSK solution at the mid-point to explain the same measurement. 

The minimum separation  is defined as the distance where the average recovery error due to the coalescence effect is equal to half the distance between each pair and their joint center of mass, at a given SNR and $M{/}N$. Inaccuracy in the recovery due to this limitation can be addressed by adjusting $M{/}N$ given by the real-space resampling, to be compatible with the estimated measured SNR. The minimal separation distance acts as the de-facto  resolution limit for an unknown sparse charge density distribution for the case of super-resolution in noisy conditions. Applying to the recovered density an additional real-space blur of the same width can be used to ensure the objectivity of the result for cases where the sparse density might contain features below the minimum separation. The implementation of this approach for general anisotropic signals can be found in the supplementary material.
 
 \section{Results}

 \begin{figure}[hbt!]
	\includegraphics[width=0.5\textwidth]{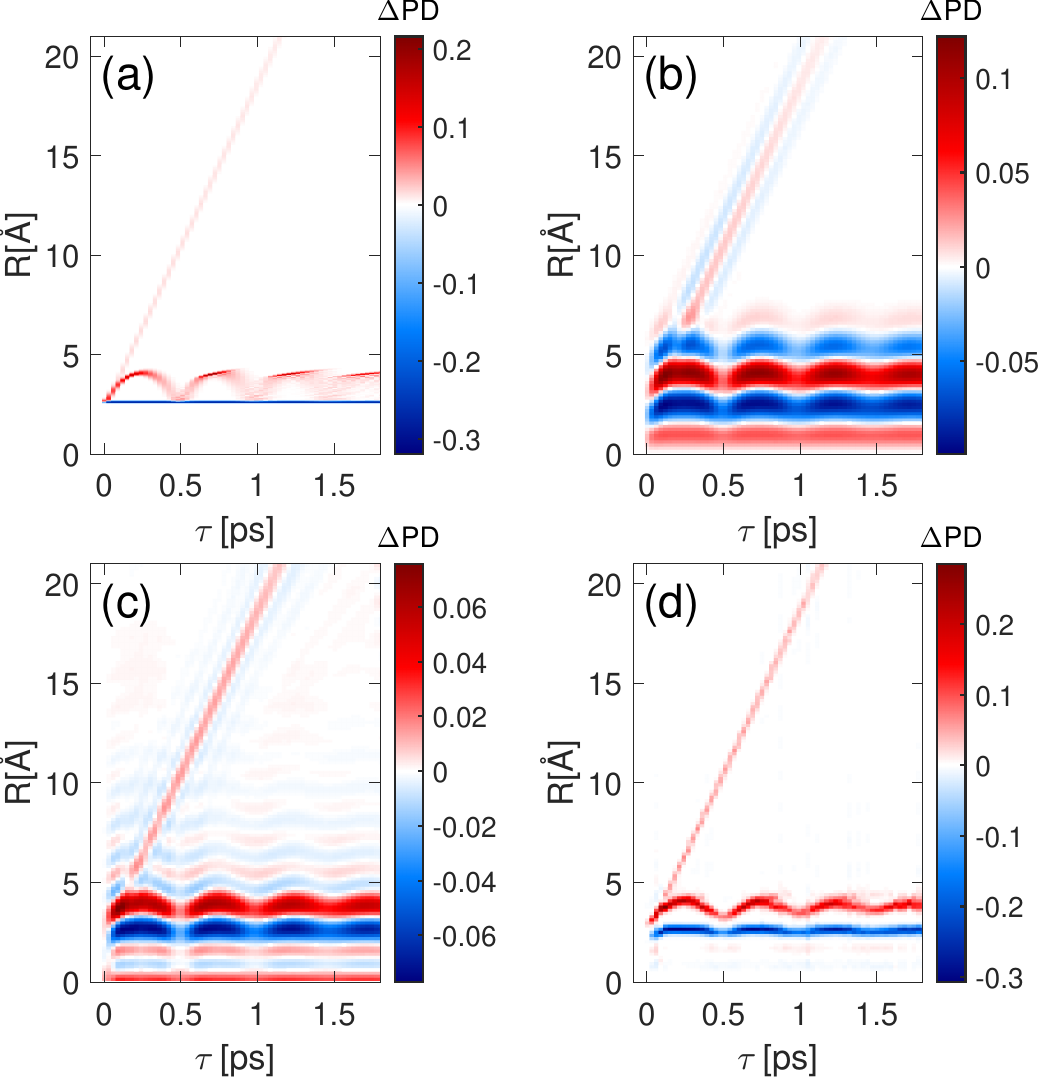} 
	\caption{(a)  TDSE calculation of the time-dependent charge density difference ($PD$ normalized to the total charge) for diatomic Iodine resulting in coherent vibration and dissociation wavepacket motions. Simulating  its  difference scattering signal assuming typical experimental conditions similar to \cite{PhysRevLett.117.153003} and adding noise (20 db SNR) we (c) inverted the truncated signal difference and obtained the pair-density difference signal. The limited $q$- range hinders the ability to recover the positions and motions of the calculated charge density.  The NSK dictionary formed by the measurement constraints was used to deconvolve the distorted pair density using (c)  $\ell_2$ and (d) $\ell_1$ regularization. For the $\ell_1$ case we obtain super-resolution in the recovery, resolving multiple motions and wavepacket shape dispersion.}
	\label{fig:NSK}
\end{figure} 

 \subsection{Numerical demonstration: Wavepacket dynamics in Iodine}

We revisit the numerical example shown in Fig \ref{fig:dicts} and consider a broader range of time-delays $0{<}\tau{<}1.8$ ps,  capturing the evolving coherent nuclear  wave-packet dynamics that undergo  multiple motions such as dissociation and unharmonic vibrations.  We simulated the scattering signal with the same measurement constraints of the previous example ($0.5{<}q{<}4\angstrom^{\text{-}1}$), and assumed an SNR of 20 dB per detector pixel on the scattering difference signal $\Delta S_0(q,\tau)$.   We inverted the truncated signal to obtain $\Delta PD_0(R,\tau)$ and used the dictionary of NSKs with the same measurement constraints to deconvolve the distorted  $\Delta PD_0(R,\tau)$ using the $\ell_2$ regularization, as seen in Fig \ref{fig:NSK}c.  While this approach handles some of the inversion artifacts, the nature of the smooth solutions it promotes reduces the ability to resolve weaker signals and does not achieve super-resolution.

In Fig \ref{fig:NSK}d, we applied convex optimization for the $\ell_1$ regularization case (Eq.\ref{eq:Regu}), apllied the minimal separation found in the previous section, and managed to  accurately capture real-space information and to super-resolve details below 0.4 $\angstrom$. For example, we were able to resolve the wavepacket positions and dispersion surpassing the diffraction limited result seen in the case of the $\ell_2$ regularization.

 \subsection{Experimental demonstration: Ring opening dynamics in CHD }

 \begin{figure}[hbt!]
	\includegraphics[width=0.5\textwidth]{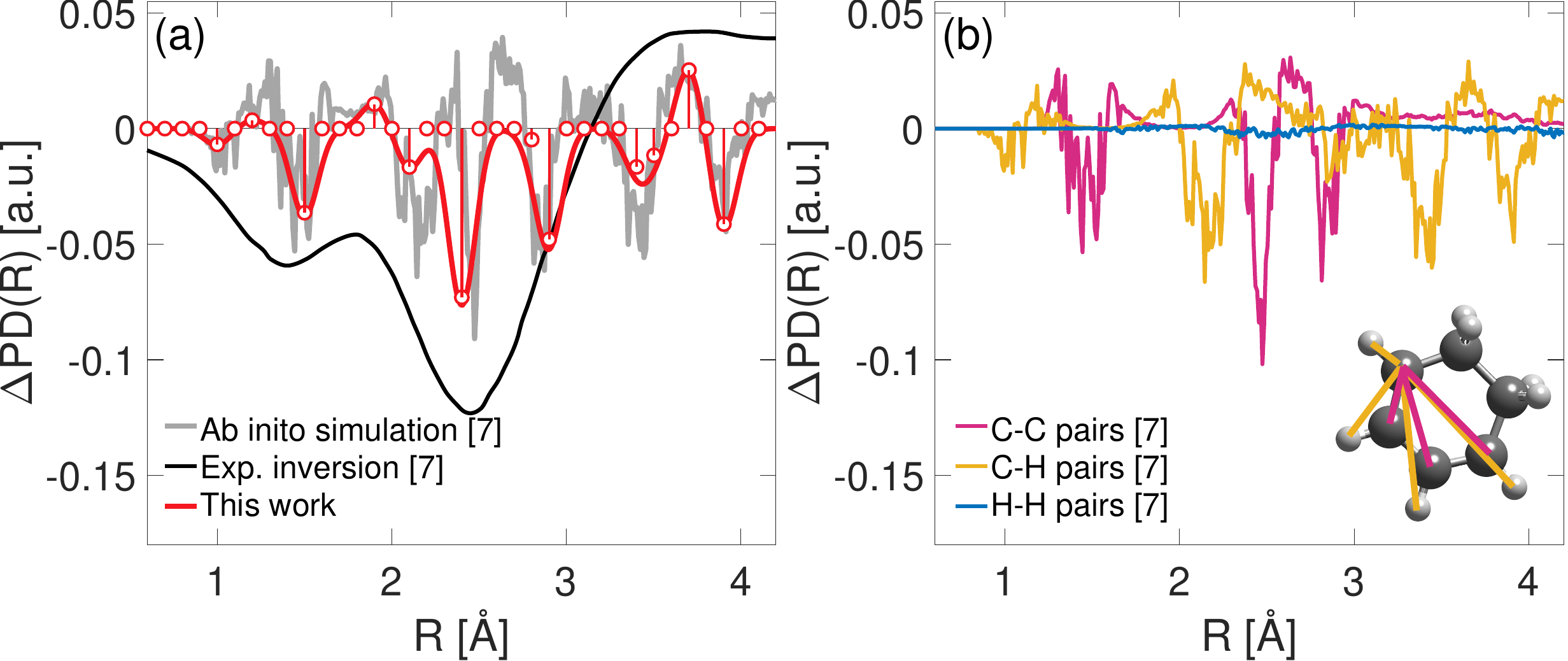} 
	\caption{ (a) Ab Initio Multiple Spawning trajectory simulation (gray) and experimental (black) pair density difference of photoinduced ring-opening of CHD at 0.55 ps from \cite{wolf2019photochemical} vs (red) the inversion and super-resolution of the experimental data using the NSKs method presented here. The NSKs solution uses $\ell_1$ regularization (red circles), following the minimal separation assumption given the estimated measurement noise (solid red).
 Individual peaks, resolved at ${<}0.3 \angstrom$ resolution correspond to atom pair distances which significantly change during the ring-opening: positive peaks point to new distances that are formed as the ring opens, while negative peaks point to depletion of steady state distances. The details resolved  
surpass the diffraction limit (${\simeq} 0.62 \angstrom$) for the measurement by a factor of ${>}$2. 
 (b) The  simulated charge density difference is obtained by including all distances and averaging over all trajectories.  Its composition to the different groups of pairs highlights the  depletion of the C-C (pink) and C-H (yellow) distances that are resolved by the NSK method. These distances  are illustrated in the CHD steady-state geometry (inset).  }

  	\label{fig:chd}
\end{figure} 

In Fig \ref{fig:chd}, we applied the NSK inversion approach using the $\ell_1$ regularization to data collected by an ultrafast electron scattering experiment that captured the photoinduced ring-opening of 1,3-cyclohexadiene (CHD) \cite{wolf2019photochemical}. This well-studied polyatomic molecule has 14 atoms and a total of 91 atom pairs in three groups (C-C, C-H, H-H).  We used the effective form factor approximation to invert the experimental signal at 0.55 ps delay,  as the structural opening of the ring takes place and 1,3,5-hexatriene (HT) isomers are formed. We analyzed the scattering signal with the same binning as in the inversion of \cite{wolf2019photochemical}, $\Delta q{=}0.0974 \angstrom^{\text{-}1}$, q-range truncation $1.3{<}q{<}10.2 \angstrom^{\text{-}1}$, and used NSK spatial resolution of $\Delta R{\simeq}0.1 \angstrom$, which corresponds to a super-resolution factor $M/N{\simeq}3.1$.  We estimated  -16 db SNR per pixel of the measured difference data by comparing the theoretical simulated scattering and measured signals, and inferred a minimum separation distance of $\delta R {\simeq} 0.15 \angstrom$ by performing a noise analysis procedure similar to the one discussed in the previous section using the q-range reported in the experiment. The minimal separation distance was then used to blur the $\ell_1$ regularization solution seen in Fig \ref{fig:chd}a.

We're able to resolve ${<}0.3\angstrom$ features, including details of several individual pair distances that were not resolved by the inversion done in \cite{wolf2019photochemical}.  Our results are in excellent agreement with the ab initio multiple spawning (AIMS) simulations, done at the $\alpha$-CASSCF(6,4)/6–31G$^*$ level of theory without further processing the trajectories in the same way as the experimental inversion done in \cite{wolf2019photochemical}. We observe the depletion of the initial steady-state C-C distances at 1.4, 2.5, 2.85 $\angstrom$, and C-H distance at 1.05, 2.1 $\angstrom$  as new distances ${>}3\angstrom$ form due to the ring opening. The new distances correspond to C-C and C-H pairs of the HT isomers. we find that the C-H related distances significantly contribute and are better markers for structural dynamics of the ring opening process due to the rigidity of the C-H bond. The details of the inversion we perform facilitate a quantitative understanding of the different contributions of the different HT isomers and help elucidate structure beyond 3 $\angstrom$ that cannot be attributed to a combination of the isomer structures.

\section{Conclusion}
 In summary, we have developed a model-free approach to directly resolve scattering signals in real space, beyond the diffraction limit.  We validate the approach using simulated and experimental scattering data. We demonstrate super-resolution of simultaneous motions de-novo and discuss the resolution limits as a function of fidelity.
 
This approach opens the way to directly trace the spatiotemporal shape of coherent wavepacket motions and energy redistribution of different atom-pairs that take place simultaneously, without bias toward the Franck-Condon active modes or the constraints of normal mode analysis. This approach can be further extended considering different regularization schemes such as the elastic net \cite{zou2005regularization}, off-the-grid type methods \cite{catala2019low}, and the addition of temporal dependence of the regularizer to further constrain the dynamics. The approach also allows the inclusion of other aspects of the scattering process, such as resonant scattering, inelastic and coherent scattering cross terms, or self-scattering as in the case of laser-induced electron diffraction \cite{Dixit_Vendrell_Santra_2012,popova2015imaging,kowalewski2017monitoring,popova2018theory,simmermacher2019electronic,simmermacher2019theory,hermann2020probing,sanchez2021molecular}. 


The  approach is particularly advantageous for high-fidelity scattering signals that are expected from high-repetition rate instruments that are becoming available, and may help bridge the established pair-distribution function analysis that requires much higher $q$ ranges ($ {>}30 \angstrom^{\text{-}1}$) with time-resolved high-energy ($15{-}25 keV$) X-ray scattering, and electron scattering experiments, which just begun recording high fidelity transient signals in the ($q {\sim} 10 \angstrom^\text{{-}1}$) range.

We thank T.J.A. Wolf,  D. M. Sanchez, and T. J. Mart\'{i}nez for the experimental data and ab-initio trajectory simulations in  \cite{wolf2019photochemical}. 
This work was supported by the U.S. Department of Energy, Office of Science, Basic Energy Sciences, Chemical Sciences, Geosciences, and Biosciences Division.

\bibliography{main}

\begin{thebibliography}{68}%
\makeatletter
\providecommand \@ifxundefined [1]{%
 \@ifx{#1\undefined}
}%
\providecommand \@ifnum [1]{%
 \ifnum #1\expandafter \@firstoftwo
 \else \expandafter \@secondoftwo
 \fi
}%
\providecommand \@ifx [1]{%
 \ifx #1\expandafter \@firstoftwo
 \else \expandafter \@secondoftwo
 \fi
}%
\providecommand \natexlab [1]{#1}%
\providecommand \enquote  [1]{``#1''}%
\providecommand \bibnamefont  [1]{#1}%
\providecommand \bibfnamefont [1]{#1}%
\providecommand \citenamefont [1]{#1}%
\providecommand \href@noop [0]{\@secondoftwo}%
\providecommand \href [0]{\begingroup \@sanitize@url \@href}%
\providecommand \@href[1]{\@@startlink{#1}\@@href}%
\providecommand \@@href[1]{\endgroup#1\@@endlink}%
\providecommand \@sanitize@url [0]{\catcode `\\12\catcode `\$12\catcode
  `\&12\catcode `\#12\catcode `\^12\catcode `\_12\catcode `\%12\relax}%
\providecommand \@@startlink[1]{}%
\providecommand \@@endlink[0]{}%
\providecommand \url  [0]{\begingroup\@sanitize@url \@url }%
\providecommand \@url [1]{\endgroup\@href {#1}{\urlprefix }}%
\providecommand \urlprefix  [0]{URL }%
\providecommand \Eprint [0]{\href }%
\providecommand \doibase [0]{https://doi.org/}%
\providecommand \selectlanguage [0]{\@gobble}%
\providecommand \bibinfo  [0]{\@secondoftwo}%
\providecommand \bibfield  [0]{\@secondoftwo}%
\providecommand \translation [1]{[#1]}%
\providecommand \BibitemOpen [0]{}%
\providecommand \bibitemStop [0]{}%
\providecommand \bibitemNoStop [0]{.\EOS\space}%
\providecommand \EOS [0]{\spacefactor3000\relax}%
\providecommand \BibitemShut  [1]{\csname bibitem#1\endcsname}%
\let\auto@bib@innerbib\@empty
\bibitem [{\citenamefont {K{\"u}pper}\ \emph {et~al.}(2014)\citenamefont
  {K{\"u}pper}, \citenamefont {Stern}, \citenamefont {Holmegaard},
  \citenamefont {Filsinger}, \citenamefont {Rouz{\'e}e}, \citenamefont
  {Rudenko}, \citenamefont {Johnsson}, \citenamefont {Martin}, \citenamefont
  {Adolph}, \citenamefont {Aquila} \emph {et~al.}}]{kupper2014x}%
  \BibitemOpen
  \bibfield  {author} {\bibinfo {author} {\bibfnamefont {J.}~\bibnamefont
  {K{\"u}pper}}, \bibinfo {author} {\bibfnamefont {S.}~\bibnamefont {Stern}},
  \bibinfo {author} {\bibfnamefont {L.}~\bibnamefont {Holmegaard}}, \bibinfo
  {author} {\bibfnamefont {F.}~\bibnamefont {Filsinger}}, \bibinfo {author}
  {\bibfnamefont {A.}~\bibnamefont {Rouz{\'e}e}}, \bibinfo {author}
  {\bibfnamefont {A.}~\bibnamefont {Rudenko}}, \bibinfo {author} {\bibfnamefont
  {P.}~\bibnamefont {Johnsson}}, \bibinfo {author} {\bibfnamefont {A.~V.}\
  \bibnamefont {Martin}}, \bibinfo {author} {\bibfnamefont {M.}~\bibnamefont
  {Adolph}}, \bibinfo {author} {\bibfnamefont {A.}~\bibnamefont {Aquila}},
  \emph {et~al.},\ }\bibfield  {title} {\bibinfo {title} {X-ray diffraction
  from isolated and strongly aligned gas-phase molecules with a free-electron
  laser},\ }\href@noop {} {\bibfield  {journal} {\bibinfo  {journal} {Physical
  Review Letters}\ }\textbf {\bibinfo {volume} {112}},\ \bibinfo {pages}
  {083002} (\bibinfo {year} {2014})}\BibitemShut {NoStop}%
\bibitem [{\citenamefont {Minitti}\ \emph
  {et~al.}(2015{\natexlab{a}})\citenamefont {Minitti}, \citenamefont {Budarz},
  \citenamefont {Kirrander}, \citenamefont {Robinson}, \citenamefont {Ratner},
  \citenamefont {Lane}, \citenamefont {Zhu}, \citenamefont {Glownia},
  \citenamefont {Kozina}, \citenamefont {Lemke}, \citenamefont {Sikorski},
  \citenamefont {Feng}, \citenamefont {Nelson}, \citenamefont {Saita},
  \citenamefont {Stankus}, \citenamefont {Northey}, \citenamefont {Hastings},\
  and\ \citenamefont {Weber}}]{PhysRevLett.114.255501}%
  \BibitemOpen
  \bibfield  {author} {\bibinfo {author} {\bibfnamefont {M.~P.}\ \bibnamefont
  {Minitti}}, \bibinfo {author} {\bibfnamefont {J.~M.}\ \bibnamefont {Budarz}},
  \bibinfo {author} {\bibfnamefont {A.}~\bibnamefont {Kirrander}}, \bibinfo
  {author} {\bibfnamefont {J.~S.}\ \bibnamefont {Robinson}}, \bibinfo {author}
  {\bibfnamefont {D.}~\bibnamefont {Ratner}}, \bibinfo {author} {\bibfnamefont
  {T.~J.}\ \bibnamefont {Lane}}, \bibinfo {author} {\bibfnamefont
  {D.}~\bibnamefont {Zhu}}, \bibinfo {author} {\bibfnamefont {J.~M.}\
  \bibnamefont {Glownia}}, \bibinfo {author} {\bibfnamefont {M.}~\bibnamefont
  {Kozina}}, \bibinfo {author} {\bibfnamefont {H.~T.}\ \bibnamefont {Lemke}},
  \bibinfo {author} {\bibfnamefont {M.}~\bibnamefont {Sikorski}}, \bibinfo
  {author} {\bibfnamefont {Y.}~\bibnamefont {Feng}}, \bibinfo {author}
  {\bibfnamefont {S.}~\bibnamefont {Nelson}}, \bibinfo {author} {\bibfnamefont
  {K.}~\bibnamefont {Saita}}, \bibinfo {author} {\bibfnamefont
  {B.}~\bibnamefont {Stankus}}, \bibinfo {author} {\bibfnamefont
  {T.}~\bibnamefont {Northey}}, \bibinfo {author} {\bibfnamefont {J.~B.}\
  \bibnamefont {Hastings}},\ and\ \bibinfo {author} {\bibfnamefont {P.~M.}\
  \bibnamefont {Weber}},\ }\bibfield  {title} {\bibinfo {title} {Imaging
  molecular motion: Femtosecond x-ray scattering of an electrocyclic chemical
  reaction},\ }\href {https://doi.org/10.1103/PhysRevLett.114.255501}
  {\bibfield  {journal} {\bibinfo  {journal} {Phys. Rev. Lett.}\ }\textbf
  {\bibinfo {volume} {114}},\ \bibinfo {pages} {255501} (\bibinfo {year}
  {2015}{\natexlab{a}})}\BibitemShut {NoStop}%
\bibitem [{\citenamefont {Glownia}\ \emph {et~al.}(2016)\citenamefont
  {Glownia}, \citenamefont {Natan}, \citenamefont {Cryan}, \citenamefont
  {Hartsock}, \citenamefont {Kozina}, \citenamefont {Minitti}, \citenamefont
  {Nelson}, \citenamefont {Robinson}, \citenamefont {Sato}, \citenamefont {van
  Driel}, \citenamefont {Welch}, \citenamefont {Weninger}, \citenamefont
  {Zhu},\ and\ \citenamefont {Bucksbaum}}]{PhysRevLett.117.153003}%
  \BibitemOpen
  \bibfield  {author} {\bibinfo {author} {\bibfnamefont {J.~M.}\ \bibnamefont
  {Glownia}}, \bibinfo {author} {\bibfnamefont {A.}~\bibnamefont {Natan}},
  \bibinfo {author} {\bibfnamefont {J.~P.}\ \bibnamefont {Cryan}}, \bibinfo
  {author} {\bibfnamefont {R.}~\bibnamefont {Hartsock}}, \bibinfo {author}
  {\bibfnamefont {M.}~\bibnamefont {Kozina}}, \bibinfo {author} {\bibfnamefont
  {M.~P.}\ \bibnamefont {Minitti}}, \bibinfo {author} {\bibfnamefont
  {S.}~\bibnamefont {Nelson}}, \bibinfo {author} {\bibfnamefont
  {J.}~\bibnamefont {Robinson}}, \bibinfo {author} {\bibfnamefont
  {T.}~\bibnamefont {Sato}}, \bibinfo {author} {\bibfnamefont {T.}~\bibnamefont
  {van Driel}}, \bibinfo {author} {\bibfnamefont {G.}~\bibnamefont {Welch}},
  \bibinfo {author} {\bibfnamefont {C.}~\bibnamefont {Weninger}}, \bibinfo
  {author} {\bibfnamefont {D.}~\bibnamefont {Zhu}},\ and\ \bibinfo {author}
  {\bibfnamefont {P.~H.}\ \bibnamefont {Bucksbaum}},\ }\bibfield  {title}
  {\bibinfo {title} {Self-referenced coherent diffraction x-ray movie of
  \aa{}ngstrom- and femtosecond-scale atomic motion},\ }\href
  {https://doi.org/10.1103/PhysRevLett.117.153003} {\bibfield  {journal}
  {\bibinfo  {journal} {Phys. Rev. Lett.}\ }\textbf {\bibinfo {volume} {117}},\
  \bibinfo {pages} {153003} (\bibinfo {year} {2016})}\BibitemShut {NoStop}%
\bibitem [{\citenamefont {Bucksbaum}\ \emph {et~al.}(2020)\citenamefont
  {Bucksbaum}, \citenamefont {Ware}, \citenamefont {Natan}, \citenamefont
  {Cryan},\ and\ \citenamefont {Glownia}}]{bucksbaum2020characterizing}%
  \BibitemOpen
  \bibfield  {author} {\bibinfo {author} {\bibfnamefont {P.~H.}\ \bibnamefont
  {Bucksbaum}}, \bibinfo {author} {\bibfnamefont {M.~R.}\ \bibnamefont {Ware}},
  \bibinfo {author} {\bibfnamefont {A.}~\bibnamefont {Natan}}, \bibinfo
  {author} {\bibfnamefont {J.~P.}\ \bibnamefont {Cryan}},\ and\ \bibinfo
  {author} {\bibfnamefont {J.~M.}\ \bibnamefont {Glownia}},\ }\bibfield
  {title} {\bibinfo {title} {Characterizing multiphoton excitation using
  time-resolved x-ray scattering},\ }\href@noop {} {\bibfield  {journal}
  {\bibinfo  {journal} {Physical Review X}\ }\textbf {\bibinfo {volume} {10}},\
  \bibinfo {pages} {011065} (\bibinfo {year} {2020})}\BibitemShut {NoStop}%
\bibitem [{\citenamefont {Natan}\ \emph {et~al.}(2021)\citenamefont {Natan},
  \citenamefont {Schori}, \citenamefont {Owolabi}, \citenamefont {Cryan},
  \citenamefont {Glownia},\ and\ \citenamefont
  {Bucksbaum}}]{natan2021resolving}%
  \BibitemOpen
  \bibfield  {author} {\bibinfo {author} {\bibfnamefont {A.}~\bibnamefont
  {Natan}}, \bibinfo {author} {\bibfnamefont {A.}~\bibnamefont {Schori}},
  \bibinfo {author} {\bibfnamefont {G.}~\bibnamefont {Owolabi}}, \bibinfo
  {author} {\bibfnamefont {J.~P.}\ \bibnamefont {Cryan}}, \bibinfo {author}
  {\bibfnamefont {J.~M.}\ \bibnamefont {Glownia}},\ and\ \bibinfo {author}
  {\bibfnamefont {P.~H.}\ \bibnamefont {Bucksbaum}},\ }\bibfield  {title}
  {\bibinfo {title} {Resolving multiphoton processes with high-order anisotropy
  ultrafast x-ray scattering},\ }\href@noop {} {\bibfield  {journal} {\bibinfo
  {journal} {Faraday Discussions}\ }\textbf {\bibinfo {volume} {228}},\
  \bibinfo {pages} {123} (\bibinfo {year} {2021})}\BibitemShut {NoStop}%
\bibitem [{\citenamefont {Kierspel}\ \emph {et~al.}(2020)\citenamefont
  {Kierspel}, \citenamefont {Morgan}, \citenamefont {Wiese}, \citenamefont
  {Mullins}, \citenamefont {Aquila}, \citenamefont {Barty}, \citenamefont
  {Bean}, \citenamefont {Boll}, \citenamefont {Boutet}, \citenamefont
  {Bucksbaum} \emph {et~al.}}]{kierspel2020x}%
  \BibitemOpen
  \bibfield  {author} {\bibinfo {author} {\bibfnamefont {T.}~\bibnamefont
  {Kierspel}}, \bibinfo {author} {\bibfnamefont {A.}~\bibnamefont {Morgan}},
  \bibinfo {author} {\bibfnamefont {J.}~\bibnamefont {Wiese}}, \bibinfo
  {author} {\bibfnamefont {T.}~\bibnamefont {Mullins}}, \bibinfo {author}
  {\bibfnamefont {A.}~\bibnamefont {Aquila}}, \bibinfo {author} {\bibfnamefont
  {A.}~\bibnamefont {Barty}}, \bibinfo {author} {\bibfnamefont
  {R.}~\bibnamefont {Bean}}, \bibinfo {author} {\bibfnamefont {R.}~\bibnamefont
  {Boll}}, \bibinfo {author} {\bibfnamefont {S.}~\bibnamefont {Boutet}},
  \bibinfo {author} {\bibfnamefont {P.}~\bibnamefont {Bucksbaum}}, \emph
  {et~al.},\ }\bibfield  {title} {\bibinfo {title} {X-ray diffractive imaging
  of controlled gas-phase molecules: Toward imaging of dynamics in the
  molecular frame},\ }\href@noop {} {\bibfield  {journal} {\bibinfo  {journal}
  {The Journal of Chemical Physics}\ }\textbf {\bibinfo {volume} {152}},\
  \bibinfo {pages} {084307} (\bibinfo {year} {2020})}\BibitemShut {NoStop}%
\bibitem [{\citenamefont {Wolf}\ \emph {et~al.}(2019)\citenamefont {Wolf},
  \citenamefont {Sanchez}, \citenamefont {Yang}, \citenamefont {Parrish},
  \citenamefont {Nunes}, \citenamefont {Centurion}, \citenamefont {Coffee},
  \citenamefont {Cryan}, \citenamefont {G{\"u}hr}, \citenamefont {Hegazy} \emph
  {et~al.}}]{wolf2019photochemical}%
  \BibitemOpen
  \bibfield  {author} {\bibinfo {author} {\bibfnamefont {T.~J.}\ \bibnamefont
  {Wolf}}, \bibinfo {author} {\bibfnamefont {D.~M.}\ \bibnamefont {Sanchez}},
  \bibinfo {author} {\bibfnamefont {J.}~\bibnamefont {Yang}}, \bibinfo {author}
  {\bibfnamefont {R.}~\bibnamefont {Parrish}}, \bibinfo {author} {\bibfnamefont
  {J.}~\bibnamefont {Nunes}}, \bibinfo {author} {\bibfnamefont
  {M.}~\bibnamefont {Centurion}}, \bibinfo {author} {\bibfnamefont
  {R.}~\bibnamefont {Coffee}}, \bibinfo {author} {\bibfnamefont
  {J.}~\bibnamefont {Cryan}}, \bibinfo {author} {\bibfnamefont
  {M.}~\bibnamefont {G{\"u}hr}}, \bibinfo {author} {\bibfnamefont
  {K.}~\bibnamefont {Hegazy}}, \emph {et~al.},\ }\bibfield  {title} {\bibinfo
  {title} {The photochemical ring-opening of 1, 3-cyclohexadiene imaged by
  ultrafast electron diffraction},\ }\href@noop {} {\bibfield  {journal}
  {\bibinfo  {journal} {Nature chemistry}\ }\textbf {\bibinfo {volume} {11}},\
  \bibinfo {pages} {504} (\bibinfo {year} {2019})}\BibitemShut {NoStop}%
\bibitem [{\citenamefont {Champenois}\ \emph {et~al.}(2021)\citenamefont
  {Champenois}, \citenamefont {Sanchez}, \citenamefont {Yang}, \citenamefont
  {Figueira~Nunes}, \citenamefont {Attar}, \citenamefont {Centurion},
  \citenamefont {Forbes}, \citenamefont {G{\"u}hr}, \citenamefont {Hegazy},
  \citenamefont {Ji} \emph {et~al.}}]{champenois2021conformer}%
  \BibitemOpen
  \bibfield  {author} {\bibinfo {author} {\bibfnamefont {E.}~\bibnamefont
  {Champenois}}, \bibinfo {author} {\bibfnamefont {D.}~\bibnamefont {Sanchez}},
  \bibinfo {author} {\bibfnamefont {J.}~\bibnamefont {Yang}}, \bibinfo {author}
  {\bibfnamefont {J.}~\bibnamefont {Figueira~Nunes}}, \bibinfo {author}
  {\bibfnamefont {A.}~\bibnamefont {Attar}}, \bibinfo {author} {\bibfnamefont
  {M.}~\bibnamefont {Centurion}}, \bibinfo {author} {\bibfnamefont
  {R.}~\bibnamefont {Forbes}}, \bibinfo {author} {\bibfnamefont
  {M.}~\bibnamefont {G{\"u}hr}}, \bibinfo {author} {\bibfnamefont
  {K.}~\bibnamefont {Hegazy}}, \bibinfo {author} {\bibfnamefont
  {F.}~\bibnamefont {Ji}}, \emph {et~al.},\ }\bibfield  {title} {\bibinfo
  {title} {Conformer-specific photochemistry imaged in real space and time},\
  }\href@noop {} {\bibfield  {journal} {\bibinfo  {journal} {Science}\ }\textbf
  {\bibinfo {volume} {374}},\ \bibinfo {pages} {178} (\bibinfo {year}
  {2021})}\BibitemShut {NoStop}%
\bibitem [{\citenamefont {Yang}\ \emph {et~al.}(2020)\citenamefont {Yang},
  \citenamefont {Zhu}, \citenamefont {F.~Nunes}, \citenamefont {Yu},
  \citenamefont {Parrish}, \citenamefont {Wolf}, \citenamefont {Centurion},
  \citenamefont {G{\"u}hr}, \citenamefont {Li}, \citenamefont {Liu} \emph
  {et~al.}}]{yang2020simultaneous}%
  \BibitemOpen
  \bibfield  {author} {\bibinfo {author} {\bibfnamefont {J.}~\bibnamefont
  {Yang}}, \bibinfo {author} {\bibfnamefont {X.}~\bibnamefont {Zhu}}, \bibinfo
  {author} {\bibfnamefont {J.~P.}\ \bibnamefont {F.~Nunes}}, \bibinfo {author}
  {\bibfnamefont {J.~K.}\ \bibnamefont {Yu}}, \bibinfo {author} {\bibfnamefont
  {R.~M.}\ \bibnamefont {Parrish}}, \bibinfo {author} {\bibfnamefont {T.~J.}\
  \bibnamefont {Wolf}}, \bibinfo {author} {\bibfnamefont {M.}~\bibnamefont
  {Centurion}}, \bibinfo {author} {\bibfnamefont {M.}~\bibnamefont {G{\"u}hr}},
  \bibinfo {author} {\bibfnamefont {R.}~\bibnamefont {Li}}, \bibinfo {author}
  {\bibfnamefont {Y.}~\bibnamefont {Liu}}, \emph {et~al.},\ }\bibfield  {title}
  {\bibinfo {title} {Simultaneous observation of nuclear and electronic
  dynamics by ultrafast electron diffraction},\ }\href@noop {} {\bibfield
  {journal} {\bibinfo  {journal} {Science}\ }\textbf {\bibinfo {volume}
  {368}},\ \bibinfo {pages} {885} (\bibinfo {year} {2020})}\BibitemShut
  {NoStop}%
\bibitem [{\citenamefont {Ihee}\ \emph {et~al.}(2010)\citenamefont {Ihee},
  \citenamefont {Wulff}, \citenamefont {Kim},\ and\ \citenamefont
  {Adachi}}]{ihee2010ultrafast}%
  \BibitemOpen
  \bibfield  {author} {\bibinfo {author} {\bibfnamefont {H.}~\bibnamefont
  {Ihee}}, \bibinfo {author} {\bibfnamefont {M.}~\bibnamefont {Wulff}},
  \bibinfo {author} {\bibfnamefont {J.}~\bibnamefont {Kim}},\ and\ \bibinfo
  {author} {\bibfnamefont {S.-i.}\ \bibnamefont {Adachi}},\ }\bibfield  {title}
  {\bibinfo {title} {Ultrafast x-ray scattering: structural dynamics from
  diatomic to protein molecules},\ }\href@noop {} {\bibfield  {journal}
  {\bibinfo  {journal} {International Reviews in Physical Chemistry}\ }\textbf
  {\bibinfo {volume} {29}},\ \bibinfo {pages} {453} (\bibinfo {year}
  {2010})}\BibitemShut {NoStop}%
\bibitem [{\citenamefont {Kim}\ \emph {et~al.}(2015)\citenamefont {Kim},
  \citenamefont {Kim}, \citenamefont {Nozawa}, \citenamefont {Sato},
  \citenamefont {Oang}, \citenamefont {Kim}, \citenamefont {Ki}, \citenamefont
  {Jo}, \citenamefont {Park}, \citenamefont {Song} \emph
  {et~al.}}]{kim2015direct}%
  \BibitemOpen
  \bibfield  {author} {\bibinfo {author} {\bibfnamefont {K.~H.}\ \bibnamefont
  {Kim}}, \bibinfo {author} {\bibfnamefont {J.~G.}\ \bibnamefont {Kim}},
  \bibinfo {author} {\bibfnamefont {S.}~\bibnamefont {Nozawa}}, \bibinfo
  {author} {\bibfnamefont {T.}~\bibnamefont {Sato}}, \bibinfo {author}
  {\bibfnamefont {K.~Y.}\ \bibnamefont {Oang}}, \bibinfo {author}
  {\bibfnamefont {T.~W.}\ \bibnamefont {Kim}}, \bibinfo {author} {\bibfnamefont
  {H.}~\bibnamefont {Ki}}, \bibinfo {author} {\bibfnamefont {J.}~\bibnamefont
  {Jo}}, \bibinfo {author} {\bibfnamefont {S.}~\bibnamefont {Park}}, \bibinfo
  {author} {\bibfnamefont {C.}~\bibnamefont {Song}}, \emph {et~al.},\
  }\bibfield  {title} {\bibinfo {title} {Direct observation of bond formation
  in solution with femtosecond x-ray scattering},\ }\href@noop {} {\bibfield
  {journal} {\bibinfo  {journal} {Nature}\ }\textbf {\bibinfo {volume} {518}},\
  \bibinfo {pages} {385} (\bibinfo {year} {2015})}\BibitemShut {NoStop}%
\bibitem [{\citenamefont {Biasin}\ \emph {et~al.}(2016)\citenamefont {Biasin},
  \citenamefont {van Driel}, \citenamefont {Kj\ae{}r}, \citenamefont {Dohn},
  \citenamefont {Christensen}, \citenamefont {Harlang}, \citenamefont {Vester},
  \citenamefont {Chabera}, \citenamefont {Liu}, \citenamefont {Uhlig},
  \citenamefont {P\'apai}, \citenamefont {N\'emeth}, \citenamefont {Hartsock},
  \citenamefont {Liang}, \citenamefont {Zhang}, \citenamefont {Alonso-Mori},
  \citenamefont {Chollet}, \citenamefont {Glownia}, \citenamefont {Nelson},
  \citenamefont {Sokaras}, \citenamefont {Assefa}, \citenamefont {Britz},
  \citenamefont {Galler}, \citenamefont {Gawelda}, \citenamefont {Bressler},
  \citenamefont {Gaffney}, \citenamefont {Lemke}, \citenamefont {M\o{}ller},
  \citenamefont {Nielsen}, \citenamefont {Sundstr\"om}, \citenamefont
  {Vank\'o}, \citenamefont {W\"arnmark}, \citenamefont {Canton},\ and\
  \citenamefont {Haldrup}}]{PhysRevLett.117.013002}%
  \BibitemOpen
  \bibfield  {author} {\bibinfo {author} {\bibfnamefont {E.}~\bibnamefont
  {Biasin}}, \bibinfo {author} {\bibfnamefont {T.~B.}\ \bibnamefont {van
  Driel}}, \bibinfo {author} {\bibfnamefont {K.~S.}\ \bibnamefont {Kj\ae{}r}},
  \bibinfo {author} {\bibfnamefont {A.~O.}\ \bibnamefont {Dohn}}, \bibinfo
  {author} {\bibfnamefont {M.}~\bibnamefont {Christensen}}, \bibinfo {author}
  {\bibfnamefont {T.}~\bibnamefont {Harlang}}, \bibinfo {author} {\bibfnamefont
  {P.}~\bibnamefont {Vester}}, \bibinfo {author} {\bibfnamefont
  {P.}~\bibnamefont {Chabera}}, \bibinfo {author} {\bibfnamefont
  {Y.}~\bibnamefont {Liu}}, \bibinfo {author} {\bibfnamefont {J.}~\bibnamefont
  {Uhlig}}, \bibinfo {author} {\bibfnamefont {M.}~\bibnamefont {P\'apai}},
  \bibinfo {author} {\bibfnamefont {Z.}~\bibnamefont {N\'emeth}}, \bibinfo
  {author} {\bibfnamefont {R.}~\bibnamefont {Hartsock}}, \bibinfo {author}
  {\bibfnamefont {W.}~\bibnamefont {Liang}}, \bibinfo {author} {\bibfnamefont
  {J.}~\bibnamefont {Zhang}}, \bibinfo {author} {\bibfnamefont
  {R.}~\bibnamefont {Alonso-Mori}}, \bibinfo {author} {\bibfnamefont
  {M.}~\bibnamefont {Chollet}}, \bibinfo {author} {\bibfnamefont {J.~M.}\
  \bibnamefont {Glownia}}, \bibinfo {author} {\bibfnamefont {S.}~\bibnamefont
  {Nelson}}, \bibinfo {author} {\bibfnamefont {D.}~\bibnamefont {Sokaras}},
  \bibinfo {author} {\bibfnamefont {T.~A.}\ \bibnamefont {Assefa}}, \bibinfo
  {author} {\bibfnamefont {A.}~\bibnamefont {Britz}}, \bibinfo {author}
  {\bibfnamefont {A.}~\bibnamefont {Galler}}, \bibinfo {author} {\bibfnamefont
  {W.}~\bibnamefont {Gawelda}}, \bibinfo {author} {\bibfnamefont
  {C.}~\bibnamefont {Bressler}}, \bibinfo {author} {\bibfnamefont {K.~J.}\
  \bibnamefont {Gaffney}}, \bibinfo {author} {\bibfnamefont {H.~T.}\
  \bibnamefont {Lemke}}, \bibinfo {author} {\bibfnamefont {K.~B.}\ \bibnamefont
  {M\o{}ller}}, \bibinfo {author} {\bibfnamefont {M.~M.}\ \bibnamefont
  {Nielsen}}, \bibinfo {author} {\bibfnamefont {V.}~\bibnamefont
  {Sundstr\"om}}, \bibinfo {author} {\bibfnamefont {G.}~\bibnamefont
  {Vank\'o}}, \bibinfo {author} {\bibfnamefont {K.}~\bibnamefont {W\"arnmark}},
  \bibinfo {author} {\bibfnamefont {S.~E.}\ \bibnamefont {Canton}},\ and\
  \bibinfo {author} {\bibfnamefont {K.}~\bibnamefont {Haldrup}},\ }\bibfield
  {title} {\bibinfo {title} {Femtosecond x-ray scattering study of ultrafast
  photoinduced structural dynamics in solvated
  $[\mathrm{Co}(\mathbf{\text{terpy}}{)}_{2}{]}^{2+}$},\ }\href
  {https://doi.org/10.1103/PhysRevLett.117.013002} {\bibfield  {journal}
  {\bibinfo  {journal} {Phys. Rev. Lett.}\ }\textbf {\bibinfo {volume} {117}},\
  \bibinfo {pages} {013002} (\bibinfo {year} {2016})}\BibitemShut {NoStop}%
\bibitem [{\citenamefont {Van~Driel}\ \emph {et~al.}(2016)\citenamefont
  {Van~Driel}, \citenamefont {Kj{\ae}r}, \citenamefont {Hartsock},
  \citenamefont {Dohn}, \citenamefont {Harlang}, \citenamefont {Chollet},
  \citenamefont {Christensen}, \citenamefont {Gawelda}, \citenamefont
  {Henriksen}, \citenamefont {Kim} \emph {et~al.}}]{van2016atomistic}%
  \BibitemOpen
  \bibfield  {author} {\bibinfo {author} {\bibfnamefont {T.~B.}\ \bibnamefont
  {Van~Driel}}, \bibinfo {author} {\bibfnamefont {K.~S.}\ \bibnamefont
  {Kj{\ae}r}}, \bibinfo {author} {\bibfnamefont {R.~W.}\ \bibnamefont
  {Hartsock}}, \bibinfo {author} {\bibfnamefont {A.~O.}\ \bibnamefont {Dohn}},
  \bibinfo {author} {\bibfnamefont {T.}~\bibnamefont {Harlang}}, \bibinfo
  {author} {\bibfnamefont {M.}~\bibnamefont {Chollet}}, \bibinfo {author}
  {\bibfnamefont {M.}~\bibnamefont {Christensen}}, \bibinfo {author}
  {\bibfnamefont {W.}~\bibnamefont {Gawelda}}, \bibinfo {author} {\bibfnamefont
  {N.~E.}\ \bibnamefont {Henriksen}}, \bibinfo {author} {\bibfnamefont {J.~G.}\
  \bibnamefont {Kim}}, \emph {et~al.},\ }\bibfield  {title} {\bibinfo {title}
  {Atomistic characterization of the active-site solvation dynamics of a model
  photocatalyst},\ }\href@noop {} {\bibfield  {journal} {\bibinfo  {journal}
  {Nature communications}\ }\textbf {\bibinfo {volume} {7}},\ \bibinfo {pages}
  {1} (\bibinfo {year} {2016})}\BibitemShut {NoStop}%
\bibitem [{\citenamefont {Chollet}\ \emph {et~al.}(2015)\citenamefont
  {Chollet}, \citenamefont {Alonso-Mori}, \citenamefont {Cammarata},
  \citenamefont {Damiani}, \citenamefont {Defever}, \citenamefont {Delor},
  \citenamefont {Feng}, \citenamefont {Glownia}, \citenamefont {Langton},
  \citenamefont {Nelson},\ and\ \citenamefont {et~al.}}]{Chollet}%
  \BibitemOpen
  \bibfield  {author} {\bibinfo {author} {\bibfnamefont {M.}~\bibnamefont
  {Chollet}}, \bibinfo {author} {\bibfnamefont {R.}~\bibnamefont
  {Alonso-Mori}}, \bibinfo {author} {\bibfnamefont {M.}~\bibnamefont
  {Cammarata}}, \bibinfo {author} {\bibfnamefont {D.}~\bibnamefont {Damiani}},
  \bibinfo {author} {\bibfnamefont {J.}~\bibnamefont {Defever}}, \bibinfo
  {author} {\bibfnamefont {J.~T.}\ \bibnamefont {Delor}}, \bibinfo {author}
  {\bibfnamefont {Y.}~\bibnamefont {Feng}}, \bibinfo {author} {\bibfnamefont
  {J.~M.}\ \bibnamefont {Glownia}}, \bibinfo {author} {\bibfnamefont {J.~B.}\
  \bibnamefont {Langton}}, \bibinfo {author} {\bibfnamefont {S.}~\bibnamefont
  {Nelson}},\ and\ \bibinfo {author} {\bibnamefont {et~al.}},\ }\bibfield
  {title} {\bibinfo {title} {The x-ray pump–probe instrument at the linac
  coherent light source},\ }\href {https://doi.org/10.1107/S1600577515005135}
  {\bibfield  {journal} {\bibinfo  {journal} {Journal of Synchrotron
  Radiation}\ }\textbf {\bibinfo {volume} {22}},\ \bibinfo {pages} {503–507}
  (\bibinfo {year} {2015})}\BibitemShut {NoStop}%
\bibitem [{\citenamefont {Haldrup}\ \emph {et~al.}(2019)\citenamefont
  {Haldrup}, \citenamefont {Levi}, \citenamefont {Biasin}, \citenamefont
  {Vester}, \citenamefont {Laursen}, \citenamefont {Beyer}, \citenamefont
  {Kj\ae{}r}, \citenamefont {Brandt~van Driel}, \citenamefont {Harlang},
  \citenamefont {Dohn}, \citenamefont {Hartsock}, \citenamefont {Nelson},
  \citenamefont {Glownia}, \citenamefont {Lemke}, \citenamefont {Christensen},
  \citenamefont {Gaffney}, \citenamefont {Henriksen}, \citenamefont
  {M\o{}ller},\ and\ \citenamefont {Nielsen}}]{PhysRevLett.122.063001}%
  \BibitemOpen
  \bibfield  {author} {\bibinfo {author} {\bibfnamefont {K.}~\bibnamefont
  {Haldrup}}, \bibinfo {author} {\bibfnamefont {G.}~\bibnamefont {Levi}},
  \bibinfo {author} {\bibfnamefont {E.}~\bibnamefont {Biasin}}, \bibinfo
  {author} {\bibfnamefont {P.}~\bibnamefont {Vester}}, \bibinfo {author}
  {\bibfnamefont {M.~G.}\ \bibnamefont {Laursen}}, \bibinfo {author}
  {\bibfnamefont {F.}~\bibnamefont {Beyer}}, \bibinfo {author} {\bibfnamefont
  {K.~S.}\ \bibnamefont {Kj\ae{}r}}, \bibinfo {author} {\bibfnamefont
  {T.}~\bibnamefont {Brandt~van Driel}}, \bibinfo {author} {\bibfnamefont
  {T.}~\bibnamefont {Harlang}}, \bibinfo {author} {\bibfnamefont {A.~O.}\
  \bibnamefont {Dohn}}, \bibinfo {author} {\bibfnamefont {R.~J.}\ \bibnamefont
  {Hartsock}}, \bibinfo {author} {\bibfnamefont {S.}~\bibnamefont {Nelson}},
  \bibinfo {author} {\bibfnamefont {J.~M.}\ \bibnamefont {Glownia}}, \bibinfo
  {author} {\bibfnamefont {H.~T.}\ \bibnamefont {Lemke}}, \bibinfo {author}
  {\bibfnamefont {M.}~\bibnamefont {Christensen}}, \bibinfo {author}
  {\bibfnamefont {K.~J.}\ \bibnamefont {Gaffney}}, \bibinfo {author}
  {\bibfnamefont {N.~E.}\ \bibnamefont {Henriksen}}, \bibinfo {author}
  {\bibfnamefont {K.~B.}\ \bibnamefont {M\o{}ller}},\ and\ \bibinfo {author}
  {\bibfnamefont {M.~M.}\ \bibnamefont {Nielsen}},\ }\bibfield  {title}
  {\bibinfo {title} {Ultrafast x-ray scattering measurements of coherent
  structural dynamics on the ground-state potential energy surface of a
  diplatinum molecule},\ }\href
  {https://doi.org/10.1103/PhysRevLett.122.063001} {\bibfield  {journal}
  {\bibinfo  {journal} {Phys. Rev. Lett.}\ }\textbf {\bibinfo {volume} {122}},\
  \bibinfo {pages} {063001} (\bibinfo {year} {2019})}\BibitemShut {NoStop}%
\bibitem [{\citenamefont {Panman}\ \emph {et~al.}(2020)\citenamefont {Panman},
  \citenamefont {Biasin}, \citenamefont {Berntsson}, \citenamefont {Hermann},
  \citenamefont {Niebling}, \citenamefont {Hughes}, \citenamefont {K\"ubel},
  \citenamefont {Atkovska}, \citenamefont {Gustavsson}, \citenamefont
  {Nimmrich}, \citenamefont {Dohn}, \citenamefont {Laursen}, \citenamefont
  {Zederkof}, \citenamefont {Honarfar}, \citenamefont {Tono}, \citenamefont
  {Katayama}, \citenamefont {Owada}, \citenamefont {van Driel}, \citenamefont
  {Kjaer}, \citenamefont {Nielsen}, \citenamefont {Davidsson}, \citenamefont
  {Uhlig}, \citenamefont {Haldrup}, \citenamefont {Hub},\ and\ \citenamefont
  {Westenhoff}}]{panman2020observing}%
  \BibitemOpen
  \bibfield  {author} {\bibinfo {author} {\bibfnamefont {M.~R.}\ \bibnamefont
  {Panman}}, \bibinfo {author} {\bibfnamefont {E.}~\bibnamefont {Biasin}},
  \bibinfo {author} {\bibfnamefont {O.}~\bibnamefont {Berntsson}}, \bibinfo
  {author} {\bibfnamefont {M.}~\bibnamefont {Hermann}}, \bibinfo {author}
  {\bibfnamefont {S.}~\bibnamefont {Niebling}}, \bibinfo {author}
  {\bibfnamefont {A.~J.}\ \bibnamefont {Hughes}}, \bibinfo {author}
  {\bibfnamefont {J.}~\bibnamefont {K\"ubel}}, \bibinfo {author} {\bibfnamefont
  {K.}~\bibnamefont {Atkovska}}, \bibinfo {author} {\bibfnamefont
  {E.}~\bibnamefont {Gustavsson}}, \bibinfo {author} {\bibfnamefont
  {A.}~\bibnamefont {Nimmrich}}, \bibinfo {author} {\bibfnamefont {A.~O.}\
  \bibnamefont {Dohn}}, \bibinfo {author} {\bibfnamefont {M.}~\bibnamefont
  {Laursen}}, \bibinfo {author} {\bibfnamefont {D.~B.}\ \bibnamefont
  {Zederkof}}, \bibinfo {author} {\bibfnamefont {A.}~\bibnamefont {Honarfar}},
  \bibinfo {author} {\bibfnamefont {K.}~\bibnamefont {Tono}}, \bibinfo {author}
  {\bibfnamefont {T.}~\bibnamefont {Katayama}}, \bibinfo {author}
  {\bibfnamefont {S.}~\bibnamefont {Owada}}, \bibinfo {author} {\bibfnamefont
  {T.~B.}\ \bibnamefont {van Driel}}, \bibinfo {author} {\bibfnamefont
  {K.}~\bibnamefont {Kjaer}}, \bibinfo {author} {\bibfnamefont {M.~M.}\
  \bibnamefont {Nielsen}}, \bibinfo {author} {\bibfnamefont {J.}~\bibnamefont
  {Davidsson}}, \bibinfo {author} {\bibfnamefont {J.}~\bibnamefont {Uhlig}},
  \bibinfo {author} {\bibfnamefont {K.}~\bibnamefont {Haldrup}}, \bibinfo
  {author} {\bibfnamefont {J.~S.}\ \bibnamefont {Hub}},\ and\ \bibinfo {author}
  {\bibfnamefont {S.}~\bibnamefont {Westenhoff}},\ }\bibfield  {title}
  {\bibinfo {title} {Observing the structural evolution in the
  photodissociation of diiodomethane with femtosecond solution x-ray
  scattering},\ }\href {https://doi.org/10.1103/PhysRevLett.125.226001}
  {\bibfield  {journal} {\bibinfo  {journal} {Phys. Rev. Lett.}\ }\textbf
  {\bibinfo {volume} {125}},\ \bibinfo {pages} {226001} (\bibinfo {year}
  {2020})}\BibitemShut {NoStop}%
\bibitem [{\citenamefont {Minitti}\ \emph
  {et~al.}(2015{\natexlab{b}})\citenamefont {Minitti}, \citenamefont {Budarz},
  \citenamefont {Kirrander}, \citenamefont {Robinson}, \citenamefont {Ratner},
  \citenamefont {Lane}, \citenamefont {Zhu}, \citenamefont {Glownia},
  \citenamefont {Kozina}, \citenamefont {Lemke} \emph
  {et~al.}}]{minitti2015imaging}%
  \BibitemOpen
  \bibfield  {author} {\bibinfo {author} {\bibfnamefont {M.}~\bibnamefont
  {Minitti}}, \bibinfo {author} {\bibfnamefont {J.}~\bibnamefont {Budarz}},
  \bibinfo {author} {\bibfnamefont {A.}~\bibnamefont {Kirrander}}, \bibinfo
  {author} {\bibfnamefont {J.}~\bibnamefont {Robinson}}, \bibinfo {author}
  {\bibfnamefont {D.}~\bibnamefont {Ratner}}, \bibinfo {author} {\bibfnamefont
  {T.}~\bibnamefont {Lane}}, \bibinfo {author} {\bibfnamefont {D.}~\bibnamefont
  {Zhu}}, \bibinfo {author} {\bibfnamefont {J.}~\bibnamefont {Glownia}},
  \bibinfo {author} {\bibfnamefont {M.}~\bibnamefont {Kozina}}, \bibinfo
  {author} {\bibfnamefont {H.}~\bibnamefont {Lemke}}, \emph {et~al.},\
  }\bibfield  {title} {\bibinfo {title} {Imaging molecular motion: Femtosecond
  x-ray scattering of an electrocyclic chemical reaction},\ }\href@noop {}
  {\bibfield  {journal} {\bibinfo  {journal} {Physical review letters}\
  }\textbf {\bibinfo {volume} {114}},\ \bibinfo {pages} {255501} (\bibinfo
  {year} {2015}{\natexlab{b}})}\BibitemShut {NoStop}%
\bibitem [{\citenamefont {Stankus}\ \emph {et~al.}(2019)\citenamefont
  {Stankus}, \citenamefont {Yong}, \citenamefont {Zotev}, \citenamefont
  {Ruddock}, \citenamefont {Bellshaw}, \citenamefont {Lane}, \citenamefont
  {Liang}, \citenamefont {Boutet}, \citenamefont {Carbajo}, \citenamefont
  {Robinson} \emph {et~al.}}]{stankus2019ultrafast}%
  \BibitemOpen
  \bibfield  {author} {\bibinfo {author} {\bibfnamefont {B.}~\bibnamefont
  {Stankus}}, \bibinfo {author} {\bibfnamefont {H.}~\bibnamefont {Yong}},
  \bibinfo {author} {\bibfnamefont {N.}~\bibnamefont {Zotev}}, \bibinfo
  {author} {\bibfnamefont {J.~M.}\ \bibnamefont {Ruddock}}, \bibinfo {author}
  {\bibfnamefont {D.}~\bibnamefont {Bellshaw}}, \bibinfo {author}
  {\bibfnamefont {T.~J.}\ \bibnamefont {Lane}}, \bibinfo {author}
  {\bibfnamefont {M.}~\bibnamefont {Liang}}, \bibinfo {author} {\bibfnamefont
  {S.}~\bibnamefont {Boutet}}, \bibinfo {author} {\bibfnamefont
  {S.}~\bibnamefont {Carbajo}}, \bibinfo {author} {\bibfnamefont {J.~S.}\
  \bibnamefont {Robinson}}, \emph {et~al.},\ }\bibfield  {title} {\bibinfo
  {title} {Ultrafast x-ray scattering reveals vibrational coherence following
  rydberg excitation},\ }\href@noop {} {\bibfield  {journal} {\bibinfo
  {journal} {Nature chemistry}\ }\textbf {\bibinfo {volume} {11}},\ \bibinfo
  {pages} {716} (\bibinfo {year} {2019})}\BibitemShut {NoStop}%
\bibitem [{\citenamefont {Hell}\ and\ \citenamefont
  {Wichmann}(1994)}]{hell1994breaking}%
  \BibitemOpen
  \bibfield  {author} {\bibinfo {author} {\bibfnamefont {S.~W.}\ \bibnamefont
  {Hell}}\ and\ \bibinfo {author} {\bibfnamefont {J.}~\bibnamefont
  {Wichmann}},\ }\bibfield  {title} {\bibinfo {title} {Breaking the diffraction
  resolution limit by stimulated emission: stimulated-emission-depletion
  fluorescence microscopy},\ }\href@noop {} {\bibfield  {journal} {\bibinfo
  {journal} {Optics letters}\ }\textbf {\bibinfo {volume} {19}},\ \bibinfo
  {pages} {780} (\bibinfo {year} {1994})}\BibitemShut {NoStop}%
\bibitem [{\citenamefont {Betzig}\ \emph {et~al.}(2006)\citenamefont {Betzig},
  \citenamefont {Patterson}, \citenamefont {Sougrat}, \citenamefont
  {Lindwasser}, \citenamefont {Olenych}, \citenamefont {Bonifacino},
  \citenamefont {Davidson}, \citenamefont {Lippincott-Schwartz},\ and\
  \citenamefont {Hess}}]{betzig2006imaging}%
  \BibitemOpen
  \bibfield  {author} {\bibinfo {author} {\bibfnamefont {E.}~\bibnamefont
  {Betzig}}, \bibinfo {author} {\bibfnamefont {G.~H.}\ \bibnamefont
  {Patterson}}, \bibinfo {author} {\bibfnamefont {R.}~\bibnamefont {Sougrat}},
  \bibinfo {author} {\bibfnamefont {O.~W.}\ \bibnamefont {Lindwasser}},
  \bibinfo {author} {\bibfnamefont {S.}~\bibnamefont {Olenych}}, \bibinfo
  {author} {\bibfnamefont {J.~S.}\ \bibnamefont {Bonifacino}}, \bibinfo
  {author} {\bibfnamefont {M.~W.}\ \bibnamefont {Davidson}}, \bibinfo {author}
  {\bibfnamefont {J.}~\bibnamefont {Lippincott-Schwartz}},\ and\ \bibinfo
  {author} {\bibfnamefont {H.~F.}\ \bibnamefont {Hess}},\ }\bibfield  {title}
  {\bibinfo {title} {Imaging intracellular fluorescent proteins at nanometer
  resolution},\ }\href@noop {} {\bibfield  {journal} {\bibinfo  {journal}
  {science}\ }\textbf {\bibinfo {volume} {313}},\ \bibinfo {pages} {1642}
  (\bibinfo {year} {2006})}\BibitemShut {NoStop}%
\bibitem [{\citenamefont {Sigal}\ \emph {et~al.}(2018)\citenamefont {Sigal},
  \citenamefont {Zhou},\ and\ \citenamefont {Zhuang}}]{sigal2018visualizing}%
  \BibitemOpen
  \bibfield  {author} {\bibinfo {author} {\bibfnamefont {Y.~M.}\ \bibnamefont
  {Sigal}}, \bibinfo {author} {\bibfnamefont {R.}~\bibnamefont {Zhou}},\ and\
  \bibinfo {author} {\bibfnamefont {X.}~\bibnamefont {Zhuang}},\ }\bibfield
  {title} {\bibinfo {title} {Visualizing and discovering cellular structures
  with super-resolution microscopy},\ }\href@noop {} {\bibfield  {journal}
  {\bibinfo  {journal} {Science}\ }\textbf {\bibinfo {volume} {361}},\ \bibinfo
  {pages} {880} (\bibinfo {year} {2018})}\BibitemShut {NoStop}%
\bibitem [{\citenamefont {Phillips}\ \emph {et~al.}(2020)\citenamefont
  {Phillips}, \citenamefont {Harkiolaki}, \citenamefont {Pinto}, \citenamefont
  {Parton}, \citenamefont {Palanca}, \citenamefont {Garcia-Moreno},
  \citenamefont {Kounatidis}, \citenamefont {Sedat}, \citenamefont {Stuart},
  \citenamefont {Castello} \emph {et~al.}}]{phillips2020cryosim}%
  \BibitemOpen
  \bibfield  {author} {\bibinfo {author} {\bibfnamefont {M.~A.}\ \bibnamefont
  {Phillips}}, \bibinfo {author} {\bibfnamefont {M.}~\bibnamefont
  {Harkiolaki}}, \bibinfo {author} {\bibfnamefont {D.~M.~S.}\ \bibnamefont
  {Pinto}}, \bibinfo {author} {\bibfnamefont {R.~M.}\ \bibnamefont {Parton}},
  \bibinfo {author} {\bibfnamefont {A.}~\bibnamefont {Palanca}}, \bibinfo
  {author} {\bibfnamefont {M.}~\bibnamefont {Garcia-Moreno}}, \bibinfo {author}
  {\bibfnamefont {I.}~\bibnamefont {Kounatidis}}, \bibinfo {author}
  {\bibfnamefont {J.~W.}\ \bibnamefont {Sedat}}, \bibinfo {author}
  {\bibfnamefont {D.~I.}\ \bibnamefont {Stuart}}, \bibinfo {author}
  {\bibfnamefont {A.}~\bibnamefont {Castello}}, \emph {et~al.},\ }\bibfield
  {title} {\bibinfo {title} {Cryosim: super-resolution 3d structured
  illumination cryogenic fluorescence microscopy for correlated ultrastructural
  imaging},\ }\href@noop {} {\bibfield  {journal} {\bibinfo  {journal}
  {Optica}\ }\textbf {\bibinfo {volume} {7}},\ \bibinfo {pages} {802} (\bibinfo
  {year} {2020})}\BibitemShut {NoStop}%
\bibitem [{\citenamefont {Sroda}\ \emph {et~al.}(2020)\citenamefont {Sroda},
  \citenamefont {Makowski}, \citenamefont {Tenne}, \citenamefont {Rossman},
  \citenamefont {Lubin}, \citenamefont {Oron},\ and\ \citenamefont
  {Lapkiewicz}}]{sroda2020sofism}%
  \BibitemOpen
  \bibfield  {author} {\bibinfo {author} {\bibfnamefont {A.}~\bibnamefont
  {Sroda}}, \bibinfo {author} {\bibfnamefont {A.}~\bibnamefont {Makowski}},
  \bibinfo {author} {\bibfnamefont {R.}~\bibnamefont {Tenne}}, \bibinfo
  {author} {\bibfnamefont {U.}~\bibnamefont {Rossman}}, \bibinfo {author}
  {\bibfnamefont {G.}~\bibnamefont {Lubin}}, \bibinfo {author} {\bibfnamefont
  {D.}~\bibnamefont {Oron}},\ and\ \bibinfo {author} {\bibfnamefont
  {R.}~\bibnamefont {Lapkiewicz}},\ }\bibfield  {title} {\bibinfo {title}
  {Sofism: Super-resolution optical fluctuation image scanning microscopy},\
  }\href@noop {} {\bibfield  {journal} {\bibinfo  {journal} {Optica}\ }\textbf
  {\bibinfo {volume} {7}},\ \bibinfo {pages} {1308} (\bibinfo {year}
  {2020})}\BibitemShut {NoStop}%
\bibitem [{\citenamefont {Lorenz}\ \emph
  {et~al.}(2010{\natexlab{a}})\citenamefont {Lorenz}, \citenamefont
  {M{\o}ller},\ and\ \citenamefont {Henriksen}}]{Lorenz_Moller_Henriksen_2010}%
  \BibitemOpen
  \bibfield  {author} {\bibinfo {author} {\bibfnamefont {U.}~\bibnamefont
  {Lorenz}}, \bibinfo {author} {\bibfnamefont {K.~B.}\ \bibnamefont
  {M{\o}ller}},\ and\ \bibinfo {author} {\bibfnamefont {N.~E.}\ \bibnamefont
  {Henriksen}},\ }\bibfield  {title} {\bibinfo {title} {Theory of time-resolved
  inelastic x-ray diffraction},\ }\href
  {https://doi.org/10.1103/PhysRevA.81.023422} {\bibfield  {journal} {\bibinfo
  {journal} {Phys. Rev. A}\ }\textbf {\bibinfo {volume} {81}},\ \bibinfo
  {pages} {023422} (\bibinfo {year} {2010}{\natexlab{a}})}\BibitemShut
  {NoStop}%
\bibitem [{\citenamefont {Simmermacher}\ \emph
  {et~al.}(2019{\natexlab{a}})\citenamefont {Simmermacher}, \citenamefont
  {Henriksen}, \citenamefont {M\o{}ller}, \citenamefont {Moreno~Carrascosa},\
  and\ \citenamefont {Kirrander}}]{simmermacher2019electronic}%
  \BibitemOpen
  \bibfield  {author} {\bibinfo {author} {\bibfnamefont {M.}~\bibnamefont
  {Simmermacher}}, \bibinfo {author} {\bibfnamefont {N.~E.}\ \bibnamefont
  {Henriksen}}, \bibinfo {author} {\bibfnamefont {K.~B.}\ \bibnamefont
  {M\o{}ller}}, \bibinfo {author} {\bibfnamefont {A.}~\bibnamefont
  {Moreno~Carrascosa}},\ and\ \bibinfo {author} {\bibfnamefont
  {A.}~\bibnamefont {Kirrander}},\ }\bibfield  {title} {\bibinfo {title}
  {Electronic coherence in ultrafast x-ray scattering from molecular wave
  packets},\ }\href {https://doi.org/10.1103/PhysRevLett.122.073003} {\bibfield
   {journal} {\bibinfo  {journal} {Phys. Rev. Lett.}\ }\textbf {\bibinfo
  {volume} {122}},\ \bibinfo {pages} {073003} (\bibinfo {year}
  {2019}{\natexlab{a}})}\BibitemShut {NoStop}%
\bibitem [{\citenamefont {Lorenz}\ \emph
  {et~al.}(2010{\natexlab{b}})\citenamefont {Lorenz}, \citenamefont
  {M{\o}ller},\ and\ \citenamefont {Henriksen}}]{lorenz2010interpretation}%
  \BibitemOpen
  \bibfield  {author} {\bibinfo {author} {\bibfnamefont {U.}~\bibnamefont
  {Lorenz}}, \bibinfo {author} {\bibfnamefont {K.~B.}\ \bibnamefont
  {M{\o}ller}},\ and\ \bibinfo {author} {\bibfnamefont {N.~E.}\ \bibnamefont
  {Henriksen}},\ }\bibfield  {title} {\bibinfo {title} {On the interpretation
  of time-resolved anisotropic diffraction patterns},\ }\href@noop {}
  {\bibfield  {journal} {\bibinfo  {journal} {New Journal of Physics}\ }\textbf
  {\bibinfo {volume} {12}},\ \bibinfo {pages} {113022} (\bibinfo {year}
  {2010}{\natexlab{b}})}\BibitemShut {NoStop}%
\bibitem [{\citenamefont {Debye}(1915)}]{debye1915zerstreuung}%
  \BibitemOpen
  \bibfield  {author} {\bibinfo {author} {\bibfnamefont {P.}~\bibnamefont
  {Debye}},\ }\bibfield  {title} {\bibinfo {title} {Zerstreuung von
  r{\"o}ntgenstrahlen},\ }\href@noop {} {\bibfield  {journal} {\bibinfo
  {journal} {Annalen der Physik}\ }\textbf {\bibinfo {volume} {351}},\ \bibinfo
  {pages} {809} (\bibinfo {year} {1915})}\BibitemShut {NoStop}%
\bibitem [{\citenamefont {Dohn}\ \emph {et~al.}(2015)\citenamefont {Dohn},
  \citenamefont {Biasin}, \citenamefont {Haldrup}, \citenamefont {Nielsen},
  \citenamefont {Henriksen},\ and\ \citenamefont
  {M{\o}ller}}]{dohn2015calculation}%
  \BibitemOpen
  \bibfield  {author} {\bibinfo {author} {\bibfnamefont {A.~O.}\ \bibnamefont
  {Dohn}}, \bibinfo {author} {\bibfnamefont {E.}~\bibnamefont {Biasin}},
  \bibinfo {author} {\bibfnamefont {K.}~\bibnamefont {Haldrup}}, \bibinfo
  {author} {\bibfnamefont {M.~M.}\ \bibnamefont {Nielsen}}, \bibinfo {author}
  {\bibfnamefont {N.~E.}\ \bibnamefont {Henriksen}},\ and\ \bibinfo {author}
  {\bibfnamefont {K.~B.}\ \bibnamefont {M{\o}ller}},\ }\bibfield  {title}
  {\bibinfo {title} {On the calculation of x-ray scattering signals from
  pairwise radial distribution functions},\ }\href@noop {} {\bibfield
  {journal} {\bibinfo  {journal} {Journal of Physics B: Atomic, Molecular and
  Optical Physics}\ }\textbf {\bibinfo {volume} {48}},\ \bibinfo {pages}
  {244010} (\bibinfo {year} {2015})}\BibitemShut {NoStop}%
\bibitem [{\citenamefont {Zernike}\ and\ \citenamefont
  {Prins}(1927)}]{zernike1927beugung}%
  \BibitemOpen
  \bibfield  {author} {\bibinfo {author} {\bibfnamefont {F.~t.}\ \bibnamefont
  {Zernike}}\ and\ \bibinfo {author} {\bibfnamefont {J.}~\bibnamefont
  {Prins}},\ }\bibfield  {title} {\bibinfo {title} {Die beugung von
  r{\"o}ntgenstrahlen in fl{\"u}ssigkeiten als effekt der
  molek{\"u}lanordnung},\ }\href@noop {} {\bibfield  {journal} {\bibinfo
  {journal} {Zeitschrift f{\"u}r Physik A Hadrons and nuclei}\ }\textbf
  {\bibinfo {volume} {41}},\ \bibinfo {pages} {184} (\bibinfo {year}
  {1927})}\BibitemShut {NoStop}%
\bibitem [{\citenamefont {Williamson}\ and\ \citenamefont
  {Zewail}(1994)}]{williamson1994ultrafast}%
  \BibitemOpen
  \bibfield  {author} {\bibinfo {author} {\bibfnamefont {J.~C.}\ \bibnamefont
  {Williamson}}\ and\ \bibinfo {author} {\bibfnamefont {A.~H.}\ \bibnamefont
  {Zewail}},\ }\bibfield  {title} {\bibinfo {title} {Ultrafast electron
  diffraction. 4. molecular structures and coherent dynamics},\ }\href@noop {}
  {\bibfield  {journal} {\bibinfo  {journal} {The Journal of Physical
  Chemistry}\ }\textbf {\bibinfo {volume} {98}},\ \bibinfo {pages} {2766}
  (\bibinfo {year} {1994})}\BibitemShut {NoStop}%
\bibitem [{\citenamefont {Budarz}\ \emph {et~al.}(2016)\citenamefont {Budarz},
  \citenamefont {Minitti}, \citenamefont {Cofer-Shabica}, \citenamefont
  {Stankus}, \citenamefont {Kirrander}, \citenamefont {Hastings},\ and\
  \citenamefont {Weber}}]{budarz2016observation}%
  \BibitemOpen
  \bibfield  {author} {\bibinfo {author} {\bibfnamefont {J.}~\bibnamefont
  {Budarz}}, \bibinfo {author} {\bibfnamefont {M.}~\bibnamefont {Minitti}},
  \bibinfo {author} {\bibfnamefont {D.}~\bibnamefont {Cofer-Shabica}}, \bibinfo
  {author} {\bibfnamefont {B.}~\bibnamefont {Stankus}}, \bibinfo {author}
  {\bibfnamefont {A.}~\bibnamefont {Kirrander}}, \bibinfo {author}
  {\bibfnamefont {J.}~\bibnamefont {Hastings}},\ and\ \bibinfo {author}
  {\bibfnamefont {P.}~\bibnamefont {Weber}},\ }\bibfield  {title} {\bibinfo
  {title} {Observation of femtosecond molecular dynamics via pump--probe gas
  phase x-ray scattering},\ }\href@noop {} {\bibfield  {journal} {\bibinfo
  {journal} {Journal of Physics B: Atomic, Molecular and Optical Physics}\
  }\textbf {\bibinfo {volume} {49}},\ \bibinfo {pages} {034001} (\bibinfo
  {year} {2016})}\BibitemShut {NoStop}%
\bibitem [{\citenamefont {Van~Driel}\ \emph {et~al.}(2020)\citenamefont
  {Van~Driel}, \citenamefont {Nelson}, \citenamefont {Armenta}, \citenamefont
  {Blaj}, \citenamefont {Boo}, \citenamefont {Boutet}, \citenamefont {Doering},
  \citenamefont {Dragone}, \citenamefont {Hart}, \citenamefont {Haller} \emph
  {et~al.}}]{van2020epix10k}%
  \BibitemOpen
  \bibfield  {author} {\bibinfo {author} {\bibfnamefont {T.~B.}\ \bibnamefont
  {Van~Driel}}, \bibinfo {author} {\bibfnamefont {S.}~\bibnamefont {Nelson}},
  \bibinfo {author} {\bibfnamefont {R.}~\bibnamefont {Armenta}}, \bibinfo
  {author} {\bibfnamefont {G.}~\bibnamefont {Blaj}}, \bibinfo {author}
  {\bibfnamefont {S.}~\bibnamefont {Boo}}, \bibinfo {author} {\bibfnamefont
  {S.}~\bibnamefont {Boutet}}, \bibinfo {author} {\bibfnamefont
  {D.}~\bibnamefont {Doering}}, \bibinfo {author} {\bibfnamefont
  {A.}~\bibnamefont {Dragone}}, \bibinfo {author} {\bibfnamefont
  {P.}~\bibnamefont {Hart}}, \bibinfo {author} {\bibfnamefont {G.}~\bibnamefont
  {Haller}}, \emph {et~al.},\ }\bibfield  {title} {\bibinfo {title} {The
  epix10k 2-megapixel hard x-ray detector at lcls},\ }\href@noop {} {\bibfield
  {journal} {\bibinfo  {journal} {Journal of synchrotron radiation}\ }\textbf
  {\bibinfo {volume} {27}} (\bibinfo {year} {2020})}\BibitemShut {NoStop}%
\bibitem [{\citenamefont {Lorch}(1969)}]{lorch1969neutron}%
  \BibitemOpen
  \bibfield  {author} {\bibinfo {author} {\bibfnamefont {E.}~\bibnamefont
  {Lorch}},\ }\bibfield  {title} {\bibinfo {title} {Neutron diffraction by
  germania, silica and radiation-damaged silica glasses},\ }\href@noop {}
  {\bibfield  {journal} {\bibinfo  {journal} {Journal of Physics C: Solid State
  Physics}\ }\textbf {\bibinfo {volume} {2}},\ \bibinfo {pages} {229} (\bibinfo
  {year} {1969})}\BibitemShut {NoStop}%
\bibitem [{\citenamefont {Baddour}\ and\ \citenamefont
  {Chouinard}(2015)}]{Baddour:15}%
  \BibitemOpen
  \bibfield  {author} {\bibinfo {author} {\bibfnamefont {N.}~\bibnamefont
  {Baddour}}\ and\ \bibinfo {author} {\bibfnamefont {U.}~\bibnamefont
  {Chouinard}},\ }\bibfield  {title} {\bibinfo {title} {Theory and operational
  rules for the discrete hankel transform},\ }\href
  {https://doi.org/10.1364/JOSAA.32.000611} {\bibfield  {journal} {\bibinfo
  {journal} {J. Opt. Soc. Am. A}\ }\textbf {\bibinfo {volume} {32}},\ \bibinfo
  {pages} {611} (\bibinfo {year} {2015})}\BibitemShut {NoStop}%
\bibitem [{\citenamefont {Shannon}(1949)}]{shannon1949communication}%
  \BibitemOpen
  \bibfield  {author} {\bibinfo {author} {\bibfnamefont {C.~E.}\ \bibnamefont
  {Shannon}},\ }\bibfield  {title} {\bibinfo {title} {Communication in the
  presence of noise},\ }\href@noop {} {\bibfield  {journal} {\bibinfo
  {journal} {Proceedings of the IRE}\ }\textbf {\bibinfo {volume} {37}},\
  \bibinfo {pages} {10} (\bibinfo {year} {1949})}\BibitemShut {NoStop}%
\bibitem [{\citenamefont {Cand{\`e}s}\ \emph {et~al.}(2006)\citenamefont
  {Cand{\`e}s}, \citenamefont {Romberg},\ and\ \citenamefont
  {Tao}}]{candes2006robust}%
  \BibitemOpen
  \bibfield  {author} {\bibinfo {author} {\bibfnamefont {E.~J.}\ \bibnamefont
  {Cand{\`e}s}}, \bibinfo {author} {\bibfnamefont {J.}~\bibnamefont
  {Romberg}},\ and\ \bibinfo {author} {\bibfnamefont {T.}~\bibnamefont {Tao}},\
  }\bibfield  {title} {\bibinfo {title} {Robust uncertainty principles: Exact
  signal reconstruction from highly incomplete frequency information},\
  }\href@noop {} {\bibfield  {journal} {\bibinfo  {journal} {IEEE Transactions
  on information theory}\ }\textbf {\bibinfo {volume} {52}},\ \bibinfo {pages}
  {489} (\bibinfo {year} {2006})}\BibitemShut {NoStop}%
\bibitem [{\citenamefont {Hansen}\ and\ \citenamefont
  {O’Leary}(1993)}]{hansen1993use}%
  \BibitemOpen
  \bibfield  {author} {\bibinfo {author} {\bibfnamefont {P.~C.}\ \bibnamefont
  {Hansen}}\ and\ \bibinfo {author} {\bibfnamefont {D.~P.}\ \bibnamefont
  {O’Leary}},\ }\bibfield  {title} {\bibinfo {title} {The use of the l-curve
  in the regularization of discrete ill-posed problems},\ }\href@noop {}
  {\bibfield  {journal} {\bibinfo  {journal} {SIAM journal on scientific
  computing}\ }\textbf {\bibinfo {volume} {14}},\ \bibinfo {pages} {1487}
  (\bibinfo {year} {1993})}\BibitemShut {NoStop}%
\bibitem [{\citenamefont {Arlot}\ and\ \citenamefont
  {Celisse}(2010)}]{arlot2010survey}%
  \BibitemOpen
  \bibfield  {author} {\bibinfo {author} {\bibfnamefont {S.}~\bibnamefont
  {Arlot}}\ and\ \bibinfo {author} {\bibfnamefont {A.}~\bibnamefont
  {Celisse}},\ }\bibfield  {title} {\bibinfo {title} {A survey of
  cross-validation procedures for model selection},\ }\href@noop {} {\bibfield
  {journal} {\bibinfo  {journal} {Statistics surveys}\ }\textbf {\bibinfo
  {volume} {4}},\ \bibinfo {pages} {40} (\bibinfo {year} {2010})}\BibitemShut
  {NoStop}%
\bibitem [{\citenamefont {Golub}\ \emph {et~al.}(1999)\citenamefont {Golub},
  \citenamefont {Hansen},\ and\ \citenamefont {O'Leary}}]{golub1999tikhonov}%
  \BibitemOpen
  \bibfield  {author} {\bibinfo {author} {\bibfnamefont {G.~H.}\ \bibnamefont
  {Golub}}, \bibinfo {author} {\bibfnamefont {P.~C.}\ \bibnamefont {Hansen}},\
  and\ \bibinfo {author} {\bibfnamefont {D.~P.}\ \bibnamefont {O'Leary}},\
  }\bibfield  {title} {\bibinfo {title} {Tikhonov regularization and total
  least squares},\ }\href@noop {} {\bibfield  {journal} {\bibinfo  {journal}
  {SIAM journal on matrix analysis and applications}\ }\textbf {\bibinfo
  {volume} {21}},\ \bibinfo {pages} {185} (\bibinfo {year} {1999})}\BibitemShut
  {NoStop}%
\bibitem [{\citenamefont {Chen}\ and\ \citenamefont
  {Donoho}(1994)}]{chen1994basis}%
  \BibitemOpen
  \bibfield  {author} {\bibinfo {author} {\bibfnamefont {S.}~\bibnamefont
  {Chen}}\ and\ \bibinfo {author} {\bibfnamefont {D.}~\bibnamefont {Donoho}},\
  }\bibfield  {title} {\bibinfo {title} {Basis pursuit},\ }in\ \href@noop {}
  {\emph {\bibinfo {booktitle} {Proceedings of 1994 28th Asilomar Conference on
  Signals, Systems and Computers}}},\ Vol.~\bibinfo {volume} {1}\ (\bibinfo
  {organization} {IEEE},\ \bibinfo {year} {1994})\ pp.\ \bibinfo {pages}
  {41--44}\BibitemShut {NoStop}%
\bibitem [{\citenamefont {Tibshirani}(1996)}]{tibshirani1996regression}%
  \BibitemOpen
  \bibfield  {author} {\bibinfo {author} {\bibfnamefont {R.}~\bibnamefont
  {Tibshirani}},\ }\bibfield  {title} {\bibinfo {title} {Regression shrinkage
  and selection via the lasso},\ }\href@noop {} {\bibfield  {journal} {\bibinfo
   {journal} {Journal of the Royal Statistical Society: Series B
  (Methodological)}\ }\textbf {\bibinfo {volume} {58}},\ \bibinfo {pages} {267}
  (\bibinfo {year} {1996})}\BibitemShut {NoStop}%
\bibitem [{\citenamefont {Boyd}\ \emph {et~al.}(2011)\citenamefont {Boyd},
  \citenamefont {Parikh},\ and\ \citenamefont {Chu}}]{boyd2011distributed}%
  \BibitemOpen
  \bibfield  {author} {\bibinfo {author} {\bibfnamefont {S.}~\bibnamefont
  {Boyd}}, \bibinfo {author} {\bibfnamefont {N.}~\bibnamefont {Parikh}},\ and\
  \bibinfo {author} {\bibfnamefont {E.}~\bibnamefont {Chu}},\ }\href@noop {}
  {\emph {\bibinfo {title} {Distributed optimization and statistical learning
  via the alternating direction method of multipliers}}}\ (\bibinfo
  {publisher} {Now Publishers Inc},\ \bibinfo {year} {2011})\BibitemShut
  {NoStop}%
\bibitem [{\citenamefont {Parikh}\ and\ \citenamefont
  {Boyd}(2014)}]{parikh2014proximal}%
  \BibitemOpen
  \bibfield  {author} {\bibinfo {author} {\bibfnamefont {N.}~\bibnamefont
  {Parikh}}\ and\ \bibinfo {author} {\bibfnamefont {S.}~\bibnamefont {Boyd}},\
  }\bibfield  {title} {\bibinfo {title} {Proximal algorithms},\ }\href@noop {}
  {\bibfield  {journal} {\bibinfo  {journal} {Foundations and Trends in
  optimization}\ }\textbf {\bibinfo {volume} {1}},\ \bibinfo {pages} {127}
  (\bibinfo {year} {2014})}\BibitemShut {NoStop}%
\bibitem [{\citenamefont {Candes}\ and\ \citenamefont
  {Romberg}(2005)}]{candes2005l1}%
  \BibitemOpen
  \bibfield  {author} {\bibinfo {author} {\bibfnamefont {E.}~\bibnamefont
  {Candes}}\ and\ \bibinfo {author} {\bibfnamefont {J.}~\bibnamefont
  {Romberg}},\ }\bibfield  {title} {\bibinfo {title} {l1-magic: Recovery of
  sparse signals via convex programming},\ }\href@noop {} {\bibfield  {journal}
  {\bibinfo  {journal} {URL: www. acm. caltech. edu/l1magic/downloads/l1magic.
  pdf}\ }\textbf {\bibinfo {volume} {4}},\ \bibinfo {pages} {14} (\bibinfo
  {year} {2005})}\BibitemShut {NoStop}%
\bibitem [{\citenamefont {Lee}\ \emph {et~al.}(2007)\citenamefont {Lee},
  \citenamefont {Battle}, \citenamefont {Raina},\ and\ \citenamefont
  {Ng}}]{lee2007efficient}%
  \BibitemOpen
  \bibfield  {author} {\bibinfo {author} {\bibfnamefont {H.}~\bibnamefont
  {Lee}}, \bibinfo {author} {\bibfnamefont {A.}~\bibnamefont {Battle}},
  \bibinfo {author} {\bibfnamefont {R.}~\bibnamefont {Raina}},\ and\ \bibinfo
  {author} {\bibfnamefont {A.~Y.}\ \bibnamefont {Ng}},\ }\bibfield  {title}
  {\bibinfo {title} {Efficient sparse coding algorithms},\ }in\ \href@noop {}
  {\emph {\bibinfo {booktitle} {Advances in neural information processing
  systems}}}\ (\bibinfo {organization} {Citeseer},\ \bibinfo {year} {2007})\
  pp.\ \bibinfo {pages} {801--808}\BibitemShut {NoStop}%
\bibitem [{\citenamefont {Mairal}\ \emph {et~al.}(2009)\citenamefont {Mairal},
  \citenamefont {Bach}, \citenamefont {Ponce},\ and\ \citenamefont
  {Sapiro}}]{mairal2009online}%
  \BibitemOpen
  \bibfield  {author} {\bibinfo {author} {\bibfnamefont {J.}~\bibnamefont
  {Mairal}}, \bibinfo {author} {\bibfnamefont {F.}~\bibnamefont {Bach}},
  \bibinfo {author} {\bibfnamefont {J.}~\bibnamefont {Ponce}},\ and\ \bibinfo
  {author} {\bibfnamefont {G.}~\bibnamefont {Sapiro}},\ }\bibfield  {title}
  {\bibinfo {title} {Online dictionary learning for sparse coding},\ }in\
  \href@noop {} {\emph {\bibinfo {booktitle} {Proceedings of the 26th annual
  international conference on machine learning}}}\ (\bibinfo {year} {2009})\
  pp.\ \bibinfo {pages} {689--696}\BibitemShut {NoStop}%
\bibitem [{\citenamefont {Gregor}\ and\ \citenamefont
  {LeCun}(2010)}]{gregor2010learning}%
  \BibitemOpen
  \bibfield  {author} {\bibinfo {author} {\bibfnamefont {K.}~\bibnamefont
  {Gregor}}\ and\ \bibinfo {author} {\bibfnamefont {Y.}~\bibnamefont {LeCun}},\
  }\bibfield  {title} {\bibinfo {title} {Learning fast approximations of sparse
  coding},\ }in\ \href@noop {} {\emph {\bibinfo {booktitle} {Proceedings of the
  27th international conference on international conference on machine
  learning}}}\ (\bibinfo {year} {2010})\ pp.\ \bibinfo {pages}
  {399--406}\BibitemShut {NoStop}%
\bibitem [{\citenamefont {Donoho}(2006)}]{donoho2006compressed}%
  \BibitemOpen
  \bibfield  {author} {\bibinfo {author} {\bibfnamefont {D.~L.}\ \bibnamefont
  {Donoho}},\ }\bibfield  {title} {\bibinfo {title} {Compressed sensing},\
  }\href@noop {} {\bibfield  {journal} {\bibinfo  {journal} {IEEE Transactions
  on information theory}\ }\textbf {\bibinfo {volume} {52}},\ \bibinfo {pages}
  {1289} (\bibinfo {year} {2006})}\BibitemShut {NoStop}%
\bibitem [{\citenamefont {Lustig}\ \emph {et~al.}(2007)\citenamefont {Lustig},
  \citenamefont {Donoho},\ and\ \citenamefont {Pauly}}]{lustig2007sparse}%
  \BibitemOpen
  \bibfield  {author} {\bibinfo {author} {\bibfnamefont {M.}~\bibnamefont
  {Lustig}}, \bibinfo {author} {\bibfnamefont {D.}~\bibnamefont {Donoho}},\
  and\ \bibinfo {author} {\bibfnamefont {J.~M.}\ \bibnamefont {Pauly}},\
  }\bibfield  {title} {\bibinfo {title} {Sparse mri: The application of
  compressed sensing for rapid mr imaging},\ }\href@noop {} {\bibfield
  {journal} {\bibinfo  {journal} {Magnetic Resonance in Medicine: An Official
  Journal of the International Society for Magnetic Resonance in Medicine}\
  }\textbf {\bibinfo {volume} {58}},\ \bibinfo {pages} {1182} (\bibinfo {year}
  {2007})}\BibitemShut {NoStop}%
\bibitem [{\citenamefont {Duarte}\ \emph {et~al.}(2008)\citenamefont {Duarte},
  \citenamefont {Davenport}, \citenamefont {Takhar}, \citenamefont {Laska},
  \citenamefont {Sun}, \citenamefont {Kelly},\ and\ \citenamefont
  {Baraniuk}}]{duarte2008single}%
  \BibitemOpen
  \bibfield  {author} {\bibinfo {author} {\bibfnamefont {M.~F.}\ \bibnamefont
  {Duarte}}, \bibinfo {author} {\bibfnamefont {M.~A.}\ \bibnamefont
  {Davenport}}, \bibinfo {author} {\bibfnamefont {D.}~\bibnamefont {Takhar}},
  \bibinfo {author} {\bibfnamefont {J.~N.}\ \bibnamefont {Laska}}, \bibinfo
  {author} {\bibfnamefont {T.}~\bibnamefont {Sun}}, \bibinfo {author}
  {\bibfnamefont {K.~F.}\ \bibnamefont {Kelly}},\ and\ \bibinfo {author}
  {\bibfnamefont {R.~G.}\ \bibnamefont {Baraniuk}},\ }\bibfield  {title}
  {\bibinfo {title} {Single-pixel imaging via compressive sampling},\
  }\href@noop {} {\bibfield  {journal} {\bibinfo  {journal} {IEEE signal
  processing magazine}\ }\textbf {\bibinfo {volume} {25}},\ \bibinfo {pages}
  {83} (\bibinfo {year} {2008})}\BibitemShut {NoStop}%
\bibitem [{\citenamefont {Katz}\ \emph {et~al.}(2009)\citenamefont {Katz},
  \citenamefont {Bromberg},\ and\ \citenamefont
  {Silberberg}}]{katz2009compressive}%
  \BibitemOpen
  \bibfield  {author} {\bibinfo {author} {\bibfnamefont {O.}~\bibnamefont
  {Katz}}, \bibinfo {author} {\bibfnamefont {Y.}~\bibnamefont {Bromberg}},\
  and\ \bibinfo {author} {\bibfnamefont {Y.}~\bibnamefont {Silberberg}},\
  }\bibfield  {title} {\bibinfo {title} {Compressive ghost imaging},\
  }\href@noop {} {\bibfield  {journal} {\bibinfo  {journal} {Applied Physics
  Letters}\ }\textbf {\bibinfo {volume} {95}},\ \bibinfo {pages} {131110}
  (\bibinfo {year} {2009})}\BibitemShut {NoStop}%
\bibitem [{\citenamefont {Shechtman}\ \emph {et~al.}(2010)\citenamefont
  {Shechtman}, \citenamefont {Gazit}, \citenamefont {Szameit}, \citenamefont
  {Eldar},\ and\ \citenamefont {Segev}}]{shechtman2010super}%
  \BibitemOpen
  \bibfield  {author} {\bibinfo {author} {\bibfnamefont {Y.}~\bibnamefont
  {Shechtman}}, \bibinfo {author} {\bibfnamefont {S.}~\bibnamefont {Gazit}},
  \bibinfo {author} {\bibfnamefont {A.}~\bibnamefont {Szameit}}, \bibinfo
  {author} {\bibfnamefont {Y.~C.}\ \bibnamefont {Eldar}},\ and\ \bibinfo
  {author} {\bibfnamefont {M.}~\bibnamefont {Segev}},\ }\bibfield  {title}
  {\bibinfo {title} {Super-resolution and reconstruction of sparse images
  carried by incoherent light},\ }\href@noop {} {\bibfield  {journal} {\bibinfo
   {journal} {Optics letters}\ }\textbf {\bibinfo {volume} {35}},\ \bibinfo
  {pages} {1148} (\bibinfo {year} {2010})}\BibitemShut {NoStop}%
\bibitem [{\citenamefont {Solomon}\ \emph {et~al.}(2018)\citenamefont
  {Solomon}, \citenamefont {Mutzafi}, \citenamefont {Segev},\ and\
  \citenamefont {Eldar}}]{solomon2018sparsity}%
  \BibitemOpen
  \bibfield  {author} {\bibinfo {author} {\bibfnamefont {O.}~\bibnamefont
  {Solomon}}, \bibinfo {author} {\bibfnamefont {M.}~\bibnamefont {Mutzafi}},
  \bibinfo {author} {\bibfnamefont {M.}~\bibnamefont {Segev}},\ and\ \bibinfo
  {author} {\bibfnamefont {Y.~C.}\ \bibnamefont {Eldar}},\ }\bibfield  {title}
  {\bibinfo {title} {Sparsity-based super-resolution microscopy from
  correlation information},\ }\href@noop {} {\bibfield  {journal} {\bibinfo
  {journal} {Optics express}\ }\textbf {\bibinfo {volume} {26}},\ \bibinfo
  {pages} {18238} (\bibinfo {year} {2018})}\BibitemShut {NoStop}%
\bibitem [{\citenamefont {Zhang}\ \emph {et~al.}(2018)\citenamefont {Zhang},
  \citenamefont {He}, \citenamefont {Wu}, \citenamefont {Chen},\ and\
  \citenamefont {Wang}}]{zhang2018tabletop}%
  \BibitemOpen
  \bibfield  {author} {\bibinfo {author} {\bibfnamefont {A.-X.}\ \bibnamefont
  {Zhang}}, \bibinfo {author} {\bibfnamefont {Y.-H.}\ \bibnamefont {He}},
  \bibinfo {author} {\bibfnamefont {L.-A.}\ \bibnamefont {Wu}}, \bibinfo
  {author} {\bibfnamefont {L.-M.}\ \bibnamefont {Chen}},\ and\ \bibinfo
  {author} {\bibfnamefont {B.-B.}\ \bibnamefont {Wang}},\ }\bibfield  {title}
  {\bibinfo {title} {Tabletop x-ray ghost imaging with ultra-low radiation},\
  }\href@noop {} {\bibfield  {journal} {\bibinfo  {journal} {Optica}\ }\textbf
  {\bibinfo {volume} {5}},\ \bibinfo {pages} {374} (\bibinfo {year}
  {2018})}\BibitemShut {NoStop}%
\bibitem [{\citenamefont {Driver}\ \emph {et~al.}(2020)\citenamefont {Driver},
  \citenamefont {Li}, \citenamefont {Champenois}, \citenamefont {Duris},
  \citenamefont {Ratner}, \citenamefont {Lane}, \citenamefont {Rosenberger},
  \citenamefont {Al-Haddad}, \citenamefont {Averbukh}, \citenamefont {Barnard}
  \emph {et~al.}}]{driver2020attosecond}%
  \BibitemOpen
  \bibfield  {author} {\bibinfo {author} {\bibfnamefont {T.}~\bibnamefont
  {Driver}}, \bibinfo {author} {\bibfnamefont {S.}~\bibnamefont {Li}}, \bibinfo
  {author} {\bibfnamefont {E.~G.}\ \bibnamefont {Champenois}}, \bibinfo
  {author} {\bibfnamefont {J.}~\bibnamefont {Duris}}, \bibinfo {author}
  {\bibfnamefont {D.}~\bibnamefont {Ratner}}, \bibinfo {author} {\bibfnamefont
  {T.~J.}\ \bibnamefont {Lane}}, \bibinfo {author} {\bibfnamefont
  {P.}~\bibnamefont {Rosenberger}}, \bibinfo {author} {\bibfnamefont
  {A.}~\bibnamefont {Al-Haddad}}, \bibinfo {author} {\bibfnamefont
  {V.}~\bibnamefont {Averbukh}}, \bibinfo {author} {\bibfnamefont
  {T.}~\bibnamefont {Barnard}}, \emph {et~al.},\ }\bibfield  {title} {\bibinfo
  {title} {Attosecond transient absorption spooktroscopy: a ghost imaging
  approach to ultrafast absorption spectroscopy},\ }\href@noop {} {\bibfield
  {journal} {\bibinfo  {journal} {Physical Chemistry Chemical Physics}\
  }\textbf {\bibinfo {volume} {22}},\ \bibinfo {pages} {2704} (\bibinfo {year}
  {2020})}\BibitemShut {NoStop}%
\bibitem [{\citenamefont {Grant}\ and\ \citenamefont
  {Boyd}(2014)}]{grant2014cvx}%
  \BibitemOpen
  \bibfield  {author} {\bibinfo {author} {\bibfnamefont {M.}~\bibnamefont
  {Grant}}\ and\ \bibinfo {author} {\bibfnamefont {S.}~\bibnamefont {Boyd}},\
  }\href@noop {} {\bibinfo {title} {Cvx: Matlab software for disciplined convex
  programming, version 2.1}} (\bibinfo {year} {2014})\BibitemShut {NoStop}%
\bibitem [{\citenamefont {Cand{\`e}s}\ and\ \citenamefont
  {Fernandez-Granda}(2013)}]{candes2013super}%
  \BibitemOpen
  \bibfield  {author} {\bibinfo {author} {\bibfnamefont {E.~J.}\ \bibnamefont
  {Cand{\`e}s}}\ and\ \bibinfo {author} {\bibfnamefont {C.}~\bibnamefont
  {Fernandez-Granda}},\ }\bibfield  {title} {\bibinfo {title} {Super-resolution
  from noisy data},\ }\href@noop {} {\bibfield  {journal} {\bibinfo  {journal}
  {Journal of Fourier Analysis and Applications}\ }\textbf {\bibinfo {volume}
  {19}},\ \bibinfo {pages} {1229} (\bibinfo {year} {2013})}\BibitemShut
  {NoStop}%
\bibitem [{\citenamefont {Fernandez-Granda}(2016)}]{fernandez2016super}%
  \BibitemOpen
  \bibfield  {author} {\bibinfo {author} {\bibfnamefont {C.}~\bibnamefont
  {Fernandez-Granda}},\ }\bibfield  {title} {\bibinfo {title} {Super-resolution
  of point sources via convex programming},\ }\href@noop {} {\bibfield
  {journal} {\bibinfo  {journal} {Information and Inference: A Journal of the
  IMA}\ }\textbf {\bibinfo {volume} {5}},\ \bibinfo {pages} {251} (\bibinfo
  {year} {2016})}\BibitemShut {NoStop}%
\bibitem [{\citenamefont {Duval}\ and\ \citenamefont
  {Peyr{\'e}}(2015)}]{duval2015exact}%
  \BibitemOpen
  \bibfield  {author} {\bibinfo {author} {\bibfnamefont {V.}~\bibnamefont
  {Duval}}\ and\ \bibinfo {author} {\bibfnamefont {G.}~\bibnamefont
  {Peyr{\'e}}},\ }\bibfield  {title} {\bibinfo {title} {Exact support recovery
  for sparse spikes deconvolution},\ }\href@noop {} {\bibfield  {journal}
  {\bibinfo  {journal} {Foundations of Computational Mathematics}\ }\textbf
  {\bibinfo {volume} {15}},\ \bibinfo {pages} {1315} (\bibinfo {year}
  {2015})}\BibitemShut {NoStop}%
\bibitem [{\citenamefont {Zou}\ and\ \citenamefont
  {Hastie}(2005)}]{zou2005regularization}%
  \BibitemOpen
  \bibfield  {author} {\bibinfo {author} {\bibfnamefont {H.}~\bibnamefont
  {Zou}}\ and\ \bibinfo {author} {\bibfnamefont {T.}~\bibnamefont {Hastie}},\
  }\bibfield  {title} {\bibinfo {title} {Regularization and variable selection
  via the elastic net},\ }\href@noop {} {\bibfield  {journal} {\bibinfo
  {journal} {Journal of the royal statistical society: series B (statistical
  methodology)}\ }\textbf {\bibinfo {volume} {67}},\ \bibinfo {pages} {301}
  (\bibinfo {year} {2005})}\BibitemShut {NoStop}%
\bibitem [{\citenamefont {Catala}\ \emph {et~al.}(2019)\citenamefont {Catala},
  \citenamefont {Duval},\ and\ \citenamefont {Peyr{\'e}}}]{catala2019low}%
  \BibitemOpen
  \bibfield  {author} {\bibinfo {author} {\bibfnamefont {P.}~\bibnamefont
  {Catala}}, \bibinfo {author} {\bibfnamefont {V.}~\bibnamefont {Duval}},\ and\
  \bibinfo {author} {\bibfnamefont {G.}~\bibnamefont {Peyr{\'e}}},\ }\bibfield
  {title} {\bibinfo {title} {A low-rank approach to off-the-grid sparse
  superresolution},\ }\href@noop {} {\bibfield  {journal} {\bibinfo  {journal}
  {SIAM Journal on Imaging Sciences}\ }\textbf {\bibinfo {volume} {12}},\
  \bibinfo {pages} {1464} (\bibinfo {year} {2019})}\BibitemShut {NoStop}%
\bibitem [{\citenamefont {Dixit}\ \emph {et~al.}(2012)\citenamefont {Dixit},
  \citenamefont {Vendrell},\ and\ \citenamefont
  {Santra}}]{Dixit_Vendrell_Santra_2012}%
  \BibitemOpen
  \bibfield  {author} {\bibinfo {author} {\bibfnamefont {G.}~\bibnamefont
  {Dixit}}, \bibinfo {author} {\bibfnamefont {O.}~\bibnamefont {Vendrell}},\
  and\ \bibinfo {author} {\bibfnamefont {R.}~\bibnamefont {Santra}},\
  }\bibfield  {title} {\bibinfo {title} {Imaging electronic quantum motion with
  light},\ }\href {https://doi.org/10.1073/pnas.1202226109} {\bibfield
  {journal} {\bibinfo  {journal} {Proceedings of the National Academy of
  Sciences}\ }\textbf {\bibinfo {volume} {109}},\ \bibinfo {pages}
  {11636–11640} (\bibinfo {year} {2012})}\BibitemShut {NoStop}%
\bibitem [{\citenamefont {Popova-Gorelova}\ and\ \citenamefont
  {Santra}(2015)}]{popova2015imaging}%
  \BibitemOpen
  \bibfield  {author} {\bibinfo {author} {\bibfnamefont {D.}~\bibnamefont
  {Popova-Gorelova}}\ and\ \bibinfo {author} {\bibfnamefont {R.}~\bibnamefont
  {Santra}},\ }\bibfield  {title} {\bibinfo {title} {Imaging interatomic
  electron current in crystals with ultrafast resonant x-ray scattering},\
  }\href@noop {} {\bibfield  {journal} {\bibinfo  {journal} {Physical Review
  B}\ }\textbf {\bibinfo {volume} {92}},\ \bibinfo {pages} {184304} (\bibinfo
  {year} {2015})}\BibitemShut {NoStop}%
\bibitem [{\citenamefont {Kowalewski}\ \emph {et~al.}(2017)\citenamefont
  {Kowalewski}, \citenamefont {Bennett},\ and\ \citenamefont
  {Mukamel}}]{kowalewski2017monitoring}%
  \BibitemOpen
  \bibfield  {author} {\bibinfo {author} {\bibfnamefont {M.}~\bibnamefont
  {Kowalewski}}, \bibinfo {author} {\bibfnamefont {K.}~\bibnamefont
  {Bennett}},\ and\ \bibinfo {author} {\bibfnamefont {S.}~\bibnamefont
  {Mukamel}},\ }\bibfield  {title} {\bibinfo {title} {Monitoring nonadiabatic
  avoided crossing dynamics in molecules by ultrafast x-ray diffraction},\
  }\href@noop {} {\bibfield  {journal} {\bibinfo  {journal} {Structural
  Dynamics}\ }\textbf {\bibinfo {volume} {4}},\ \bibinfo {pages} {054101}
  (\bibinfo {year} {2017})}\BibitemShut {NoStop}%
\bibitem [{\citenamefont {Popova-Gorelova}\ \emph {et~al.}(2018)\citenamefont
  {Popova-Gorelova}, \citenamefont {Reis},\ and\ \citenamefont
  {Santra}}]{popova2018theory}%
  \BibitemOpen
  \bibfield  {author} {\bibinfo {author} {\bibfnamefont {D.}~\bibnamefont
  {Popova-Gorelova}}, \bibinfo {author} {\bibfnamefont {D.~A.}\ \bibnamefont
  {Reis}},\ and\ \bibinfo {author} {\bibfnamefont {R.}~\bibnamefont {Santra}},\
  }\bibfield  {title} {\bibinfo {title} {Theory of x-ray scattering from
  laser-driven electronic systems},\ }\href@noop {} {\bibfield  {journal}
  {\bibinfo  {journal} {Physical Review B}\ }\textbf {\bibinfo {volume} {98}},\
  \bibinfo {pages} {224302} (\bibinfo {year} {2018})}\BibitemShut {NoStop}%
\bibitem [{\citenamefont {Simmermacher}\ \emph
  {et~al.}(2019{\natexlab{b}})\citenamefont {Simmermacher}, \citenamefont
  {Moreno~Carrascosa}, \citenamefont {E.~Henriksen}, \citenamefont
  {B.~M{\o}ller},\ and\ \citenamefont {Kirrander}}]{simmermacher2019theory}%
  \BibitemOpen
  \bibfield  {author} {\bibinfo {author} {\bibfnamefont {M.}~\bibnamefont
  {Simmermacher}}, \bibinfo {author} {\bibfnamefont {A.}~\bibnamefont
  {Moreno~Carrascosa}}, \bibinfo {author} {\bibfnamefont {N.}~\bibnamefont
  {E.~Henriksen}}, \bibinfo {author} {\bibfnamefont {K.}~\bibnamefont
  {B.~M{\o}ller}},\ and\ \bibinfo {author} {\bibfnamefont {A.}~\bibnamefont
  {Kirrander}},\ }\bibfield  {title} {\bibinfo {title} {Theory of ultrafast
  x-ray scattering by molecules in the gas phase},\ }\href@noop {} {\bibfield
  {journal} {\bibinfo  {journal} {The Journal of chemical physics}\ }\textbf
  {\bibinfo {volume} {151}},\ \bibinfo {pages} {174302} (\bibinfo {year}
  {2019}{\natexlab{b}})}\BibitemShut {NoStop}%
\bibitem [{\citenamefont {Hermann}\ \emph {et~al.}(2020)\citenamefont
  {Hermann}, \citenamefont {Pohl}, \citenamefont {Dixit},\ and\ \citenamefont
  {Tremblay}}]{hermann2020probing}%
  \BibitemOpen
  \bibfield  {author} {\bibinfo {author} {\bibfnamefont {G.}~\bibnamefont
  {Hermann}}, \bibinfo {author} {\bibfnamefont {V.}~\bibnamefont {Pohl}},
  \bibinfo {author} {\bibfnamefont {G.}~\bibnamefont {Dixit}},\ and\ \bibinfo
  {author} {\bibfnamefont {J.~C.}\ \bibnamefont {Tremblay}},\ }\bibfield
  {title} {\bibinfo {title} {Probing electronic fluxes via time-resolved x-ray
  scattering},\ }\href@noop {} {\bibfield  {journal} {\bibinfo  {journal}
  {Physical Review Letters}\ }\textbf {\bibinfo {volume} {124}},\ \bibinfo
  {pages} {013002} (\bibinfo {year} {2020})}\BibitemShut {NoStop}%
\bibitem [{\citenamefont {S{\'a}nchez}\ \emph {et~al.}(2021)\citenamefont
  {S{\'a}nchez}, \citenamefont {Amini}, \citenamefont {Wang}, \citenamefont
  {Steinle}, \citenamefont {Belsa}, \citenamefont {Danek}, \citenamefont {Le},
  \citenamefont {Liu}, \citenamefont {Moshammer}, \citenamefont {Pfeifer} \emph
  {et~al.}}]{sanchez2021molecular}%
  \BibitemOpen
  \bibfield  {author} {\bibinfo {author} {\bibfnamefont {A.}~\bibnamefont
  {S{\'a}nchez}}, \bibinfo {author} {\bibfnamefont {K.}~\bibnamefont {Amini}},
  \bibinfo {author} {\bibfnamefont {S.-J.}\ \bibnamefont {Wang}}, \bibinfo
  {author} {\bibfnamefont {T.}~\bibnamefont {Steinle}}, \bibinfo {author}
  {\bibfnamefont {B.}~\bibnamefont {Belsa}}, \bibinfo {author} {\bibfnamefont
  {J.}~\bibnamefont {Danek}}, \bibinfo {author} {\bibfnamefont {A.-T.}\
  \bibnamefont {Le}}, \bibinfo {author} {\bibfnamefont {X.}~\bibnamefont
  {Liu}}, \bibinfo {author} {\bibfnamefont {R.}~\bibnamefont {Moshammer}},
  \bibinfo {author} {\bibfnamefont {T.}~\bibnamefont {Pfeifer}}, \emph
  {et~al.},\ }\bibfield  {title} {\bibinfo {title} {Molecular structure
  retrieval directly from laboratory-frame photoelectron spectra in
  laser-induced electron diffraction},\ }\href@noop {} {\bibfield  {journal}
  {\bibinfo  {journal} {Nature communications}\ }\textbf {\bibinfo {volume}
  {12}},\ \bibinfo {pages} {1} (\bibinfo {year} {2021})}\BibitemShut {NoStop}%
\end{thebibliography}%

\end{document}


\title{Supplemental Material: Real-Space Inversion and Super-Resolution of Ultrafast Scattering}

\author{Adi Natan}
 \email{natan@stanford.edu}
\affiliation{Stanford PULSE Institute, SLAC National Accelerator Laboratory 2575 Sand Hill Road, Menlo Park, CA 94025}

\maketitle

\section{Scattering and inversion for anisotropic charge densities}
  In ultrafast scattering studies, the dynamics of a photoexcited ensemble of molecules are usually captured in a pump-probe scheme, where the excitation pump pulse is usually a linearly polarized ultrafast optical pulse with a duration shorter than the typical timescales of motion of interest.  Integrating the angle information of the scattering signal is well justified as it captures all types of motions that take place in the photoexcited system, however,  often there is an inherent anisotropy in the scattering signal when a sample is excited by linearly polarized light due to an optically induced dipole moment transition. The photo-absorption process creates charge density anisotropy in the ensemble, that can be used to filter and enhance the specific processes under study, such as in the case of a single-photon absorption process \cite{PhysRevLett.117.153003,biasin2018anisotropy,haldrup2019ultrafast}, as well as excitation via two or higher number of photons   \cite{natan2021resolving, yang2018imaging}.

We first discuss the general real-space anisotropy information of  photoexcited  molecular systems and derive the corresponding anisotropic scattering curves $S_n(q,\tau)$ using results that have been derived elsewhere \cite{Lorenz_Moller_Henriksen_2010,lorenz2010interpretation,biasin2018anisotropy,natan2021resolving}. 
We assume that both the X-ray pulse and optical laser pulse are co-propagating along $\vec k_0$, and that the angle between the incident X-ray beam and the laser polarization is $\pi{/}2$, which allows employing a Legendre decomposition in angle space. The expression for the differential cross-section for scattering is :  
   \begin{gather}
      \begin{aligned}
\frac{d\sigma}{d\Omega} &= \sigma_T   \int d \mathbf{R} \rho (\mathbf{R},\tau) |F(\mathbf{q},\mathbf{R})|^2 , \\
F(\mathbf{q},\mathbf{R})&= \sum_{b\ne a}  f_a^*(q) f_b(q) e^{\imath \mathbf{q} \mathbf{R}},
     \end{aligned}
   \end{gather}

where $\sigma_T$ is the Thomson cross-section, $\rho$ is the charge density,  and the squared molecular form factor $|F|^2$ is invariant under space inversion $R \mapsto -R$, with $f_i(q)$ the i$^{th}$ atomic form factor,  the scattering vector $\mathbf {q}{=}\mathbf{k_s}{-}\mathbf{k_0}$ and the (double) sum is over all atom pairs.  We can expand the scattering exponential term using the plane wave expansion:
\begin{equation}
e^{\imath \mathbf{q}  \mathbf{R}} = \sum_{n=0,2,\ldots} (2n+1)(-1)^{n/2} P_n(cos \theta_{qR}) j_n(qR),
\end{equation}
 
where $P_n$ are Legendre polynomials, $j_n$ are spherical Bessel functions, and $\theta_{qR}$ is the angle between $\mathbf{q}$ 
and $\mathbf{R}$.  We  sum only over even orders as odd polynomial components are anti-symmetric under space inversion and will cancel for the differential cross-section.
Using the spherical harmonics addition theorem, we expand the Legendre polynomials to express $\theta_{qR}$ in terms of the experimentally measured scattering angles $(\theta_q,\phi_q$),  and $(\theta,\phi)$, the angle between the laser polarization and dipole transition axis and its corresponding azimuth:

\begin{equation}
P_n(cos \theta_{qR}) = \frac{4 \pi}{2n+1}\sum\limits_{j=-k}^k Y_{jk}^*(\theta_q,\phi_q) Y_{jk}(\theta,\phi).
\end{equation}

We integrate over $\phi_q$ using scattering symmetry and use the above expressions to arrive to:

\begin{equation}
\frac{d\sigma}{d\Omega} = 4 \pi \sigma_T  \sum_{n=0,2,\ldots}  (-1)^{n/2} P_{n}(\cos \theta_{q}) S_n(q) ,
\label{eq:diffraction_pattern}
\end{equation}

\begin{equation}
\begin{split}
S_n(q,\tau) =   \sum_{a} f_a(q)^2+  \sum_{b\ne a} f_a^*(q) f_b(q) \int_{-1}^{1} d \cos(\theta)   \int_{0}^{\infty} dR       R^2   
\rho_{a,b}(R,\cos(\theta),\tau) P_{n}(\cos(\theta)) j_n(qR) ,
\end{split}
\label{eq:Sn}
\end{equation}

where $\rho_{a,b}(R,\cos(\theta),\tau)$ is the time-dependent and angle-dependent pair density  at time-delay $\tau$. Scattering from a general anisotropic real-space charge will be manifested both by the intensity distribution on the detector via the scattering angle $\theta_q$ and via the anisotropy curves $S_n(q)$.  In order to recover the anisotropic charge density, we need to obtain  $S_n(q)$ from the scattering pattern on the detector. This can be done by applying a Legendre decomposition over the detector angle $\theta_d$, described in \cite{natan2021resolving}.

We decompose the 2D measured detector signal in each $[q,\theta_d]$ bin to even order Legendre basis up to the relevant significant order:
\begin{equation}
I(q,\theta_d) =   \beta_0 (q) \sum_{n=0,2,\ldots} \widetilde{\beta}_n(q)  P_{n}(\cos \theta_{d})   ,
\label{eq:Leg_decomp}
\end{equation}

where $\beta_0(q)$ is  the radial detector intensity  and  $\widetilde{\beta}_n(q) = \beta_n(q) / \beta_0(q)$ are the normalized detector anisotropy terms.  The relation between the n$^{th}$ order anisotropy curve $S_n(q)$ in Eq. \ref{eq:Sn} and the corresponding  $\beta_n(q)$ term is given by \cite{natan2021resolving}:
\begin{align}
S_n(q) =    \frac{\beta_0(q) \widetilde{\beta}_n(q)} {\cos^n(\theta_q)}=  
\frac{\beta_n(q) }  { \left(1-  \frac{q^2}{4|k_0|^2}   \right)^{n/2}}  ,  &&  n=0,2,\ldots,
\label{eq:Beta2Sn}
\end{align}
where $|k_0|$ is the length of the wave vector of the incoming X-ray beam. $\beta_0(q,\tau)$ has units of intensity or the average number of scattered photons per $q$ on the detector, whereas  the anisotropic $\widetilde{\beta}_n(q,\tau)$  terms are dimensionless and represent ratios between intensities and different angles that govern  the degree of anisotropy.

   Direct real-space inversion and formation of NSKs will be applied to $S_n$ in a similar way used for the isotropic case.

The distortions are expressed as:
   
 \begin{equation}
 \tilde{S}_n(q,\tau){=} \int_{z_0{-}d}^{z_0{+}d}  dz
  \, h_n(q(z)) \, S_{n}(q(z),\tau)+b(q(z))    .
\label{eq:distortion}
\end{equation}
 Where $q(z)$ was explained in the main study, and $h_n(q)$ is a window function that represents the detector's finite $q$-range as well as $q$-dependent signal absorption, that contains angle dependence per anisotropy order $n$,  and $b(q)$ is a possible additive background.

   The discretization scheme for each anisotropy order $n$ is given by:

 \begin{align}
   \bigl\{ q^{(n)}_{i} \bigr\}_1^N &= \frac{j_{ni}}{j_{nN}}N \Delta q ,& \Bigl\{ R^{(n)}_{m} \Bigr\}_1^M &= \frac{j_{nm}}{M \Delta q}.
  \end{align}

The distorted anisotropic difference scattering signal is discretized $\Delta  \tilde{S}_n(q^{(n)}_i,\tau)$ , and the transformation back to real-space for the $n^{th}$ order anisotropy:

  \begin{equation}
 \Delta PD_n(R^{(n)},\tau)  {=} \frac{[(M{-}1) \Delta q] ^2}{j_{n M} } \sum_{i=1}^{M{-}1}  \bm{G}^{(n)}_{mi} \frac{  \Delta \tilde{S}_n(q^{(n)}_i,\tau) q^{(n)}_i }{ f_e(q^{(n)}_i)}   , 
 \label{eq:qtoR}
\end{equation}

where $f_e$ is the effective form factor, and the transformation kernel for anisotropy order $n$ is:

\begin{align}
   [\bm{G}]^{n}_{mi}   = 2\frac{j_n \left( \frac{j_{nm}j_{ni} }{j_{nM}} \right) } {j_{nM} j_{n+1}^2(j_{ni})}  ,
 \label{eq:Gki}
\end{align}
We note that for each  anisotropy order $n$ the sampling is  based on the spherical Bessel $j_n$ roots, and as a result, different orders have slightly different sampling points.

The  NSK for anisotropy order $n$  for position $R_m$, is expressed by:

\begin{equation}
 NSK_n(R_m)  {=} \frac{ [(M{-}1) \Delta q] ^2   }{j_{n M} } \sum_{i=1}^{M{-}1}  \bm{G}^{(n)}_{mi}   \tilde{j_n}(q^{(n)}_iR^{(n)}_m) q^{(n)}_i   , 
 \label{eq:NSK_deltaS}
\end{equation}
where, 

 \begin{equation}
\tilde{j}_n(q,R^{(n)}_m){=} \int_{z_0{-}d}^{z_0{+}d}  dz
  \, h_n(q(z)) \, j_{n}(q(z)R^{(n)}_m)    .
  \end{equation}

We can now create a set of NSKs for each anisotropy order along $R$ to form a set of  dictionaries $\boldsymbol{\mathcal{D}^{(n)}}$ that will be used to approximate charge density distributions of each anisotropy order. The inversion and deconvolution are then done separately for each order $\Delta S_n(q)$ using the same regularization framework. 

 \begin{equation}
 \min_{\text{w}^{(n)}}     \big\|  PD_n -\boldsymbol{\mathcal{D}}^{(n)} \mathbf{w}^{(n)}  \big\|^2 \ + \epsilon \mathcal{R}( \mathbf{w}^{(n)} ).
 \label{eq:Regu}
\end{equation}

 Each real-space anisotropic pair density solutions $\mathbf{w}^{(n)}$ has different sampling points $R^{(n)}_m$. In order to obtain the angle-resolved pair density,  we first resample $\mathbf{w}^{(n)}$ to the isotropic sampling case $R^{(0)}_m$.  We then express the weight of each order in the real space representation $\{R,\theta\}$ by:
 
 \begin{equation}
  \mathbf{w}(R,\cos \theta) = \sum_{n=0,2,\ldots} P_n(\cos \theta) {\mathbf{\tilde w}(R)^{(n)} },
\label{eq:PD}
\end{equation}

where $ \mathbf{\tilde w}^{(n)}(R)$ is the resampled $\mathbf{w}^{(n)}$ on the isotropic real-space grid $R^{(0)}_m$, up to leading anisotropy order $n$. We demonstrate this using an example similar to what was shown for the isotropic case.

\section{Numerical example}

\begin{figure}[hbt!]
	\includegraphics[width=1\textwidth]{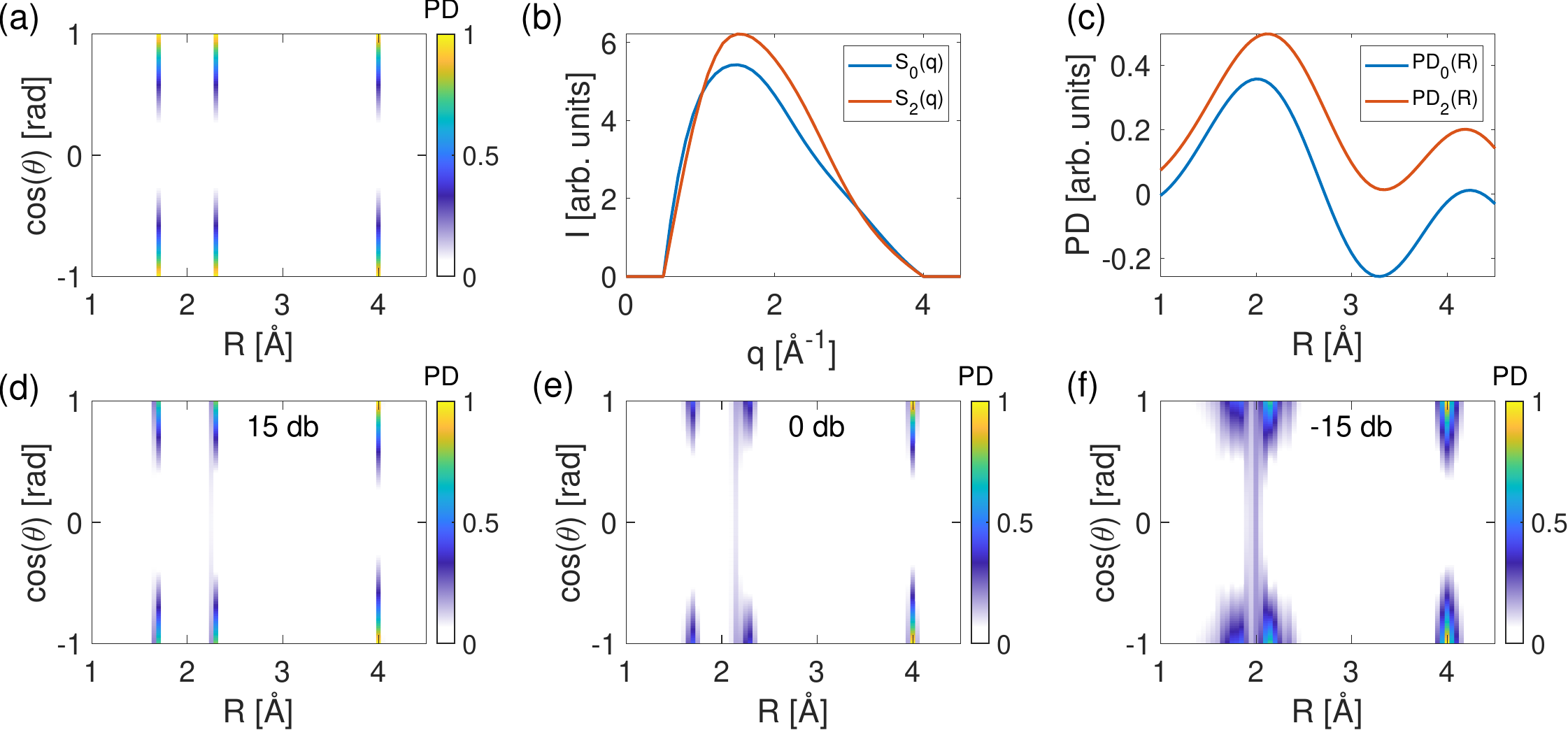} 
	\caption{Inversion and super-resolution of anisotropic charge density: (a) three pair-densities with $cos^2(\theta)$ angular distribution at $1.7\angstrom$, $2.3\angstrom$ and $4 \angstrom$ are obtained by modeling an ensemble of triatomic molecules (see text). (b) The total simulated scattering is truncated and analyzed to obtain the scattering curves $S_n$. (c) The inversion of $S_n$ fails to retrieve the positions of the pair densities. We apply the NSK inversion and deconvolution using $\ell_1$ regularization, and apply Eq \ref{eq:PD} to obtain the angle and space resolved pair densities for SNR values of (d) 15, (e) 0, and (f) -15 dB. We super-resolve distances and angular distribution and observe that the pair coalescence effect  takes place  for the  -15 dB case. At this SNR the coalescence effect limits both the spatial and angle resolution  for anisotropic scattering.}
	\label{fig:NSKani}
\end{figure}

We demonstrate the inversion of anisotropic scattering using the following example: an ensemble of linear triatomic molecules is photoexcited via absorbing a single photon, with a dipole transition parallel to the molecular axis creating a $\cos^2(\theta)$ angular distribution in the ensemble. The scattering signal from the ensemble is the incoherent sum of scattering from single molecules, as we can neglect the 2-molecule scattering signal contribution because of their random positions. Thus, we can describe the charge density signal using the ensemble-averaged molecule picture. Using the results of the isotropic case, we chose the atom positions vs the center of mass of each molecule to be for atom A at $x{=}{-}2 \angstrom$, for atom B at $x{=}0.3 \angstrom$, and for atom C at $x{=}2 \angstrom$, with the corresponding pair distances $R_{AB}{=}1.7 \angstrom$, $R_{BC}{=}2.3 \angstrom$ and $R_{AC}{=}4 \angstrom$, each with a $cos^2(\theta)$ angular distribution (Fig \ref{fig:NSKani} a).

We simulated the scattering signal on a realistic detector, assuming signal-to-noise (SNR) ratios of -15,0, and 15 dB, modeled as a $q$-dependent additive white Gaussian noise, where each $q_i$ bin has sampling statistics proportional to the number of detector pixels contributing to it.  We apply detector truncation restricting the range of measured signal to $0.5{<}q{<}4 \angstrom^{-1}$.  We note that using these parameters, the two pair distances $R_{AB}$ and $R_{BC}$ are $0.6 \angstrom$ apart, below the diffraction limit of $2\pi{/}4 \angstrom^{-1} {\sim} 1.6 \angstrom$. 

We retrieve the $S_n(q)$ curves using Eq \ref{eq:Beta2Sn} (Fig \ref{fig:NSKani} b), and invert to obtain $PD_n$ (Fig \ref{fig:NSKani} c). We form $NSK_n$ dictionaries for each $Sn$ and deconvolve $PD_n$ using Eq \ref{eq:Regu}, with $\ell_1$ regularization. The recoveries for each SNR are shown in Fig \ref{fig:NSKani} d-f, capturing both positions and angular distributions of the pair densities. We obtain similar recovery accuracy and super-resolution conditions, limited by signal fidelity and pair  coalescence. For the -15db case, the coalescence effect for angle-resolved pair distances also significantly reduces angle information related to their anisotropy. This effect can be used as a sensitive probe for the resolution limit in real-space angle-resolved recoveries.

\bibliography{supplement}
